\documentclass[journal]{IEEEtran}
\IEEEoverridecommandlockouts
\usepackage{cite}
\usepackage{amsmath, amsfonts}
\usepackage{dsfont}
\usepackage{amssymb}
\usepackage{algorithmic}
\usepackage{graphicx}
\usepackage{textcomp}
\usepackage{xcolor}
\usepackage{bbm}
\usepackage{comment}
\usepackage{enumitem}
\usepackage{makecell}
\usepackage{pgfplots}
\usepackage{textcomp}
\pgfplotsset{compat=newest}
\pgfplotsset{plot coordinates/math parser=true}
\usetikzlibrary{automata,positioning,calc,math}
\usetikzlibrary{arrows.meta}
\usetikzlibrary{arrows}
\usetikzlibrary{shapes.geometric}

\def\BibTeX{{\rm B\kern-.05em{\sc i\kern-.025em b}\kern-.08em
    T\kern-0.1667em\lower.7ex\hbox{E}\kern-0.125emX}}
    
    \usepackage{amsthm}
    \newtheorem{theorem}{Theorem}[section]

 \theoremstyle{remark}
  \newtheorem{remark}{Remark}[section]
 
\newcommand{\Id}{\mathbbm{1}} 
\newcommand{\id}{\mathrm{id}} 
\newcommand{\dd}{\ \mathrm{d}}  
\newcommand{\E}{\mathbb{E}} 
\newcommand{\V}{\mathbb{V}\mathrm{ar}} 

\newcommand{\I}{\mathbb{I}} 
\newcommand{\IR}{\overline \I}
\newcommand{\cummIR}{J}
\newcommand{\graphNullcline}{h}
\newcommand{\Cov}{\mathbb{C}\mathrm{ov}}
\newcommand{\Prob}{\mathbb{P}} 
\newcommand{\R}{\mathbb{R}} 
\newcommand{\N}{\mathbb{N}} 
 
\newcommand{\sgn}{\mathrm{sgn}} 
\newcommand{\sigmalg}{\mathcal F} 
\newcommand{\vol}{\mathrm{vol}}

\newmuskip\pFqmuskip

\newcommand*\pFq[6][8]{%
  \begingroup 
  \pFqmuskip=#1mu\relax
  \mathchardef\normalcomma=\mathcode`,
  \mathcode`\,=\string"8000
  \begingroup\lccode`\~=`\,
  \lowercase{\endgroup\let~}\pFqcomma
  F^{#2}_{#3}{\left[\genfrac..{0pt}{}{#4}{#5};#6\right]}%
  \endgroup
}
\newcommand{\pFqcomma}{{\normalcomma}\mskip\pFqmuskip}

\newcommand{\msg}{X}    
\newcommand{\msgStateSpace}{\mathcal X}    
\newcommand{\msgState}{x}    
\newcommand{\out}{Y}    
\newcommand{\num}{n}    
\newcommand{\condmean}{Z} 

\newcommand{\z}{z}
 
\newcommand{\Pois}{\mathrm{Pois}}

\newcommand{\target}{g} 
\newcommand{\jtarget}{f}
\newcommand{\auxODE}{u}
\newcommand{\CIprojection}{l}
\newcommand{\extension}{\Sigma}
\newcommand{\updateProj}{\Gamma}
\newcommand{\SuffDyn}{F}

\newcommand{\CProb}{\mathbf\Pi}
\newcommand{\cprob}{\pi}
\newcommand{\intensity}{\lambda}
\newcommand{\intensityFct}{\lambda}
\newcommand{\CIntensity}{\hat \lambda}
\newcommand{\cintensity}{m}
\newcommand{\aci}{\cintensity}
\newcommand{\acifrom}{\cintensity'}

\newcommand{\gain}{c}

\newcommand{\Density}{\rho} 
\newcommand{\density}{p}    
\newcommand{\idensity}{p_0} 
\newcommand{\idensityCoeff}{a}
\newcommand{\mdensityCoeff}{p}
\newcommand{\acid}{p_\lambda}
\newcommand{\integrand}{I}

\newcommand{\MasterEq}{\mathcal A} 
\newcommand{\OpI}{A} 
\newcommand{\semiMarkovKernel}{Q}
\newcommand{\tKernel}{K}
\newcommand{\niter}{L}
\newcommand{\aciKernel}{K} 

\newcommand{\state}{\theta}
\newcommand{\stateSet}{B} 
\newcommand{\statefrom}{\theta'} 
\newcommand{\stateto}{\theta}
\newcommand{\stateSupport}{\Omega}
\newcommand{\statePartition}{\Omega}
\newcommand{\ipart}{i} 
\newcommand{\iPart}{j} 
\newcommand{\inum}{N} 
\newcommand{\itrap}{I} 
\newcommand{\iTrap}{I'} 
\newcommand{\posSet}{\mathcal C_+}
\newcommand{\negSet}{\mathcal C_-}
\newcommand{\weight}{\nu} 
\newcommand{\path}{\gamma}
\newcommand{\low}{a}
\newcommand{\up}{b}
\newcommand{\numPath}{\eta}
\newcommand{\SuffStat}{V} 
\newcommand{\suffStat}{v} 
\newcommand{\intervalBound}{b}
\newcommand{\mintervalBound}{m}

\newcommand{\ti}{t}
\newcommand{\intime}{\tau}
\newcommand{\itime}{k}
\newcommand{\wtime}{T}

\newcommand{\HawkesKernel}{H}
\newcommand{\HawkesBase}{\mu_0}
\newcommand{\HawkesJump}{\beta}
\newcommand{\HawkesDecay}{\alpha}
\newcommand{\HawkesMean}{\mu}
\newcommand{\HawkesVariance}{\sigma^2}
\newcommand{\HawkesAutocov}{\gamma}
\newcommand{\HawkesMartingale}{Q}

\newcommand{\DonsoffOn}{\alpha_{01}}
\newcommand{\DonsoffR}{\alpha_{11}}
\newcommand{\DonsoffOff}{\alpha_{10}}
\newcommand{\DonsoffFrac}{U}
\newcommand{\Donsofffrac}{u}
\newcommand{\OnValue}{0.04}

\newcommand{\OffValue}{1.6}
\newcommand{\DonsoffMIValue}{0.101}
\newcommand{\RTMIValue}{0.096}

\newcommand{\leakage}{\lambda_0}
\newcommand{\ONState}{\lambda_1}
\newcommand{\dynamicRange}{\Delta \lambda}
\newcommand{\LeakageOn}{k_1}
\newcommand{\LeakageOff}{k_2}

\newcommand{\RTOn}{k_1}
\newcommand{\RTOff}{k_2}
\newcommand{\RTON}{\tilde{k}_1}
\newcommand{\RTOFF}{\tilde{k}_2}
\newcommand{\RTRoot}{\omega}
\newcommand{\gainSc}{\alpha}

\newcommand{\OnConstraint}{r_1}
\newcommand{\OffConstraint}{r_2}
\newcommand{\intersection}{0.29}
\newcommand{\MaxOn}{\RTOn^\ast}
\newcommand{\MaxOff}{\RTOff^\ast}
\newcommand{\energy}{c}

\newcommand{\GammaMean}{M}
\newcommand{\GammaVar}{S}
\newcommand{\Gammavar}{s}
\newcommand{\GammaMu}{\mu}
\newcommand{\GammaSigma}{\sigma^2}
\newcommand{\GammaDecay}{\gamma}

\newcommand{\gammamean}{m}

\newcommand{\PDMP}{Z} 
\newcommand{\PDMPDyn}{A} 
\newcommand{\PDMPTarget}{f} 
\newcommand{\PDMPTargInv}{f_{-}}
\newcommand{\PDMPState}{z}
\newcommand{\PDMPDens}{p}
\newcommand{\PDMPStatefrom}{z'}
\newcommand{\UpperBound}{b}
\newcommand{\PDMPHelp}{G}

\newlength{\myfigwidth}
 \usepackage{setspace}

\begin{document}
\title{ACID: A Low Dimensional Characterization of Markov-Modulated and Self-Exciting Counting Processes}

\author{Mark~Sinzger-D'Angelo and
        Heinz~Koeppl
\thanks{This paper was presented in part at the 2020 IEEE International Symposium on Information Theory, virtual, Los Angeles, California, USA.}
\thanks{The work was supported by the European Research Council (ERC) within the Consolidator Grant CONSYN (grant agreement no. 773196).}%
\thanks{M. Sinzger-D'Angelo and H. Koeppl are with the Department of Electrical Engineering and Information Technology, Technische Universität Darmstadt (email: \{mark.sinzger, heinz.koeppl\}@tu-darmstadt.de). H. Koeppl is with the Centre for Synthetic Biology, Technische Universität Darmstadt.}}
\maketitle
\begin{abstract}
The conditional intensity (CI) of a counting process $\out_\ti$ is based on the minimal knowledge $\sigmalg^\out_\ti$, i.e., on the observation of $\out_\ti$ alone. Prominently, the mutual information rate of a signal and its Poisson channel output is a difference functional between the CI and the intensity that has full knowledge about the input. While the CI of Markov-modulated Poisson processes evolves according to Snyder's filter, self-exciting processes, e.g., Hawkes processes, specify the CI via the history of $\out_\ti$. The emergence of the CI as a self-contained stochastic process prompts us to bring its statistical ensemble into focus. We investigate the asymptotic conditional intensity distribution (ACID) and emphasize its rich information content. We assume the case in which the CI is determined from a sufficient statistic that progresses as a Markov process. We present a simulation-free method to compute the ACID when the dimension of the sufficient statistic is low. The method is made possible by introducing a backward recurrence time parametrization, which has the advantage to align all probability inflow in a boundary condition for the master equation. Case studies illustrate the usage of ACID for three primary examples: 1) the Poisson channels with binary Markovian input (as an example of a Markov-modulated Poisson process), 2) the standard Hawkes process with exponential kernel (as an example of a self-exciting counting process) and 3) the Gamma filter (as an example of an approximate filter to a Markov-modulated Poisson process).
\end{abstract}

\section{Introduction}
The study of the Poisson channel dates back to the 70s. While originally introduced as a model of optical communication \cite{O.Macchi.1972}, its setting is ubiquitous in various fields of research for modeling event counts: An input signal $\msg_\ti$, possibly corrupted by dark current, is partially observed via point observations. In the terminology of Cox \cite{D.R.Cox.1955} and Snyder \cite{Snyder.1991}, the channel output $\out_\ti$ is a doubly stochastic Poisson process with feedback-less intensity $\intensity_\ti = \intensityFct(\msg_\ti)$.

Much is already known about the Poisson channel. In particular, its capacity under amplitude constraint $\msg_\ti \in [0,\energy]$ was determined by Kabanov \cite{Kabanov.1978} shortly after its introduction. However, the capacity-achieving input distribution is found to switch infinitely fast between $0$ and $\energy$ \cite{Kabanov.1978, Davis.1980}. This flaw in terms of physical interpretability was addressed by adding constraints on the bandwidth \cite{Snyder.1983,Shamai.1991,S.Shamai.1993}. While reducing the class of input processes this necessitates to review the task of computing the MI.

Besides taking its definition
    \begin{equation}
        \I(\msg_{[0,T]}, \out_{[0,T]}) := \E\left[ \ln \frac{\dd \mu^{\msg,\out}_T}{\dd \mu^\msg_T \times \mu^\out_T} \right]
        \label{def:MI}
    \end{equation}
as a departure for the computation, an expression found to be useful was obtained by Liptser \cite{Liptser.2001}
 \begin{equation}
    \I(\msg_{[0,T]}, \out_{[0,T]}) = \int_0^T \E[\phi(\intensity_\ti) - \phi(\hat \intensity_\ti)] \dd \ti,
        \label{eq:Liptser_time}
    \end{equation}
 where $\phi(z) = z \ln z$ and $\hat \intensity_\ti = \E[\intensity_\ti\vert \sigmalg^\out_\ti]$. Although other results on links between the MI and conditional estimation received considerable attention \cite{Guo.2008, Jiao.2013}, the classical expression eq.~\eqref{eq:Liptser_time} remains the standard (besides the definition eq.~\eqref{def:MI}) when searching for ways to compute the MI \cite{Lestas.2010, L.Duso.2019}.
 
 The difficulty in computing the MI via eq.~\eqref{eq:Liptser_time} clearly lies in the computation of $\E[\phi(\CIntensity_\ti)]$. Different approaches address this. (i) In case of a stationary input $\intensity_\ti$, one can consider
 $$ \intensity_\ti \to \intensity_\infty, \quad \CIntensity_\ti \to \CIntensity_\infty$$
 in distribution, still allowing conclusions about the mutual \textit{information rate}
    \begin{equation}
    \IR(\msg, \out) : = \lim_{T \to \infty}\frac 1 T \I(\msg_{[0,T]}, \out_{[0,T]}) = \E[\phi(\intensity_\infty)] - \E[\phi(\hat \intensity_\infty)].
        \label{eq:Liptser}
    \end{equation}
 (ii) Properties of the function $\phi$ (Lipschitz for bounded $\intensityFct$ \cite{Kabanov.1978} or non-negative third derivative \cite{S.Shamai.1993}) were used for a reduction to second order moments of $\CIntensity_\infty$. (iii) Kabanov, Davis used martingale theory with its rich second order analysis tools.
 (iv) The conditioning on a coarser sigma-field \cite{Shamai.1991} or the use of a suboptimal estimator \cite{S.Shamai.1993} provided upper bounds. (v) Alternatively, one needs to have the knack for choosing tractable input process classes. For example, piecewise-constant input trajectories, whose amplitudes follow a Markov chain were considered \cite{S.Shamai.1990}.

We took the approach (i) combined with (v), restricting ourselves to continuous-time Markov chains $\msg_\ti$ with low number of input states. 
    Its appearance on the right-hand side of \eqref{eq:Liptser} is our main motivation to look at the distribution of $\CIntensity_\infty$, the asymptotic conditional intensity distribution (ACID). With the spotlight being on $\CIntensity_\ti$, we elaborate on its interpretation as \textit{conditional intensity} in \ref{sub:ci}. A summary of our general contributions is followed by a confined focus on cell biology for interested readers (\ref{sub:bio}). In \ref{sub:marginal simulation} we advocate that the study of ACID can be beneficial beyond the information rate.
 
 \subsection{Conditional intensity view replaces estimator perspective}\label{sub:ci}
The conditional mean $\CIntensity_\ti = \E[\intensity_\ti \vert \sigmalg^\out_\ti]$ is often regarded as an estimator of $\intensity_\ti$. There is a clear justification for this: For each $\ti \geq 0$
    \begin{equation}
       \E[\intensity_\ti \vert \sigmalg^\out_\ti] = \arg \min_{Z} \E[(\intensity_\ti - Z)^2] 
    \end{equation}
among all random variables $Z$ that are $\sigmalg^\out_\ti$-measurable. But this interpretation attributes a notion of deficit to $\CIntensity_\ti$: While $\intensity_\ti$ is the true intensity of $\out_\ti$, we obtain merely its optimal yet inaccurate approximation via $\CIntensity_\ti$. We advocate a different perspective throughout the paper.

The process $\out_\ti$ is a \textit{self-exciting process} with intensity $\CIntensity_\ti$, which has been known at least since the monograph \cite{Snyder.1991} by Snyder. How can $\out_\ti$ have two intensities? How can $\CIntensity_\ti$ be on a par with $\intensity_\ti$?
This puzzle is solved if one incorporates that the intensity $\intensity_\ti$ of a general counting process $\out_\ti$  has an $\sigmalg_\ti$-dependency. This dependency can be expressed in at least two ways: (i) explicitly but not of operational value for mathematical proofs
$$ \intensity_\ti = \E[\dd \out_\ti\vert \sigmalg_\ti], $$
(ii) implicitly, and beneficial for mathematical rigidity \cite{Bremaud.1972} by requesting
$$ \out_\ti - \int_0^\ti \intensity_s \dd s$$
to be an $\sigmalg_\ti$-martingale.
It implies that a general counting process $\out_\ti$ can possess several intensities. For a strictly coarser $\mathcal G_\ti \subseteq \sigmalg_\ti$ and corresponding intensities $\hat \intensity_\ti, \intensity_\ti$, Bremaud's innovation theorem provides
$ \hat \intensity_\ti = \E[\intensity_\ti \vert \mathcal G_\ti] $. 
This allows the following interpretation of eq.~\eqref{eq:Liptser}. The information rate is not a difference functional between the true intensity and its deficient estimate, but between two intensities that differ in the state of knowledge: $\sigmalg^{\msg,\out}_\ti = \sigma(\msg_s,\out_s: s \le \ti)$ vs. $\sigmalg^{\out}_\ti = \sigma(\out_s: s \le \ti)$. We follow Bremaud \cite{Bremaud.1981} and Daley, Vere-Jones \cite{Daley.2003} in using the term conditional intensity for the $\sigmalg^\out_\ti$-intensity.
It is not to be confused with its other usage as a special case of the Papangelou intensity. 

Far from being an estimator with deficits, the conditional intensity $\CIntensity_\ti$ is canonical one among the $\sigmalg^{\out}_\ti$- and $\sigmalg^{\msg,\out}_\ti$-intensities. The $\sigmalg^{\out}_\ti$-intensity can be defined for any counting process $\out_\ti$, with no need to specify an external process.
And it has a standing as being the intensity with respect to the minimal filtration to which $\out_\ti$ is adapted.

The estimator perspective regards $\E[\intensity_\ti \vert \sigmalg^\out_\ti]$ as a function of the history $\{ \out_s \colon 0 \le s \le \ti \}$. This 
complicated functional dependence makes it hard
to understand $\CIntensity_\ti \to \CIntensity_\infty$. In contrast, the conditional intensity view offers a self-contained description of $\CIntensity_\ti$. It emancipates from $\out_\ti$ when both have common jumps. This view guides our understanding of $\CIntensity_\ti \to \CIntensity_\infty$.

We restrict our study of ACID to processes $\out_\ti$ for which $\CIntensity_\ti$ is obtained from a piecewise-deterministic Markov process. Within the scope of this work, we review that Markov-modulated Poisson processes belong to this class \cite{Parag.2017}, as do the Hawkes process \cite{HAWKES.1971, Oakes.1975} and approximate filters, introduced in section \ref{sub:marginal simulation}. We exploit that the ACID can then be obtained from generator theory of Markov processes\cite{Gardiner.2009}. To this end, we contribute a backward recurrence time parametrization (BReT-P), whose advantage is the alignment of all probability inflow. Since our numerical method works via grid discretization, it is limited in the dimension of the piecewise-deterministic Markov process. But conceptually, BReT-P opens the door for new numerical methods that compute the MI without Monte Carlo simulation.
We illustrate how our method helps with screening for system parameters that optimize the information rate under bandwidth-like constraints. Applications beyond information theory complement the case studies.

\subsection{Path mutual information in cell signaling}\label{sub:bio}
Our particular interest lies in studying the mutual information between time-varying signals in biological systems. We review the state of the art of information theory in cell biology to locate our contribution to the field. Readers mainly interested in the method can skip this paragraph.

It is widely assumed that the optimization of the MI could be an evolutionary strategy to mitigate noise in the cellular signalling via metabolites and in gene regulatory networks \cite{Crisanti.2018}. There is mounting evidence that information is encoded in the temporal profile of biomolecules \cite{Purvis.2013,  Selimkhanov.2014, Friedrich.2019}, but the exact mechanisms how temporal features result in decision making are unclear \cite{Maity.2020}. A path MI that is sensitive to temporal effects has been recently introduced for a class of chemical reaction networks \cite{L.Duso.2019}. The MI is often interpreted as a measure for the amount of input states that can be resolved accurately \cite{TomaszJetka.2019}. From an engineering perspective maximizing the path MI between sensor and actuator could thus define a design principle to construct cellular circuits. Information theory has been used to reveal fundamental physical limits of information transmission guiding our understanding of gene regulatory motifs as information processing units \cite{Lestas.2010, Suderman.2018}.

\begin{figure*}
    \centering
    \includegraphics[width = 0.95\textwidth]{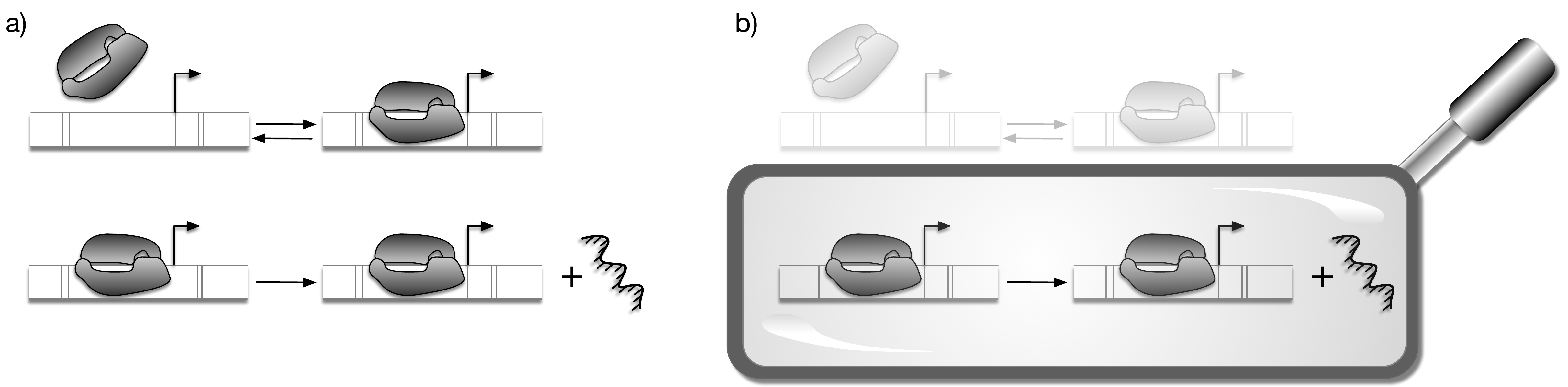}
    \caption{\textbf{The standard transcription model with a two-state promoter.} a) A random telegraph promoter switches between its active and its inactive state. Only in the active state, mRNA is synthesized. The promoter stays active, so multiple mRNA strands can be transcribed in one active period. The transcription reaction is of first order with fixed rate. b) Experimentally only the transcription events are observed. The promoter is modelled as a context. An observation model for transcription counts is obtained when we marginalize the joint system over the context. The transcription reaction is of order zero with stochastic rate, i.e., a doubly stochastic Poisson process.}
    \label{fig:Transcription_model}
\end{figure*}
Why is the Poisson channel used as a communication model in cells? Chemical reaction networks are standard Markov approaches in the stochastic modeling of gene regulation in cells. Reactants $X_1, \dots, X_n$ are converted, synthesized and degraded in reactions
$$ R_j: \sum_{i = 1}^n\nu_{ij} X_i \overset{r_j}{\longrightarrow} \sum_{i = 1}^n\eta_{ij} X_i, \quad j = 1, \dots, m. $$
This induces a continuous time Markov chain (CTMC) on the state space $\N_{\geq 0}^n$ of molecule counts $\vec{k} = (k_1, \dots, k_n)$, where the non-diagonal entry $(\vec{k} \to \vec{l})$ in the generator matrix is
$$\sum_{j = 1}^m r_j\prod_{i = 1}^n\binom{k_i}{\nu_{ij}}\mathds{1}(\vec{l}-\vec{k} = \vec{\eta}_j - \vec{\nu}_j).$$
The order of reaction $R_j$ is $\sum_{i = 1}^n \nu_{ij}$. For instance, the standard transcription model with the two-state promoter (Fig.~\ref{fig:Transcription_model}a) contains a first-order synthesis reaction
$$ P_{\mathrm{on}} \to  P_{\mathrm{on}} + \mathrm{mRNA}.$$

The homogeneous Poisson process corresponds to a zero-order reaction. First-order reactions can be interpreted as zero-order reactions with stochastic modulation, i.e., as Poisson channel with linear $\intensityFct(\msg_\ti)$ (Fig.~\ref{fig:Transcription_model}b).
More generally, the Poisson channel accounts for the discrete nature of reactions $R_j$ with the reaction counts $\out_j(\ti)$ up to time $\ti$ being the Poisson channel output. Suited for the low copy number regime, it can bring to attention the bottleneck character of sensor molecules that sense a continuous signal but whose synthesis events are restricted to the discrete regime \cite{Lestas.2010, Parag.2019}. Adhering to our goal of including temporal effects, the Poisson channel is suited because it is capable of modeling time-varying inputs and outputs.

Computing the path MI over Poissonian channels is difficult. Many works of research resort to basing the MI on single time-point marginals ignoring any encoding in the temporal profile \cite{Tkacik.2008,Suderman.2017,Crisanti.2018}.
Other approaches are the following: (i) Gaussian approximations of the input to make use of analytic results \cite{Lestas.2010}, (ii) Monte Carlo estimators \cite{Pasha.2012,CepedaHumerez.2019,L.Duso.2019}. (iii) Not rarely, the intractability decoyed researchers to resort to other channels such as the Gaussian channel \cite{Tostevin.2009,Roy.2021}. We contribute to information theory in biology by presenting a Monte Carlo-free numerical computation of the path mutual information rate of a Poisson channel for Markovian input with low state number.

\subsection{Approximate filters and approximate marginal simulation}\label{sub:marginal simulation}
The study of ACID can be beneficial beyond its appearance in the information rate of the Poisson channel.
Markov-modulated Poisson processes do not only serve as communication channel models. They provide observation models for open systems, i.e., the counting process can be regarded as a subsystem that is embedded in a heat bath \cite{Breuer.2002} or environment \cite{Zechner.2014}. The Markovian environment modulates the intensity of the observed subsystem.

In the biological cell, for instance,  the rate of transcription synthesis events is modulated by multiple factors \cite{ZhixingCao.2020}. Recruitment of polymerase and transcription factors as well as unwinding of the DNA strand, all contribute to the activation of a promoter state, where transcription is initiated \cite{Lenstra.2016}. The combination of these factors can be regarded as a random environmental process modulating the rates in the observed main process, here transcription synthesis (Fig.~\ref{fig:Transcription_model}b).

An observation model of the subsystem requires modeling the unobserved context, e.g., as Markovian environment.
After marginalizing over the environment, the CI describes the main process in an uncoupled way.
Sample trajectories, i.e., Monte Carlo samples, can be obtained by simulating the main process as a self-exciting counting process whose rate is the CI. In practice, this is done via thinning \cite{P.A.WLewis.1979} or by the inverse transform method to sample sojourn times.
This \textit{marginal simulation} replaces the co-simulation of the environment and promises a potential speed-up. Only the \textit{effective} rate $\CIntensity_\ti$ is used, neglecting fluctuations of the environmental process that are not transmitted to the main process. However, unclosed conditional moment equations pose a serious challenge to achieving the goal of fast marginal simulation \cite{Zechner.2014}. This problem is addressed by approximate filters obtained from conditional moment closure, assumed density filtering, variational inference, entropic matching or projection \cite{Zechner.2016, Bronstein.2018, Snyder.1991}. Prominently, the Hawkes process can be obtained from projection onto the class of linear estimators and can thus be regarded as an approximate filter. We call the marginal simulation with an approximate filter \textit{approximate marginal simulation}.

The process that results from approximate marginal simulation is self-exciting, but in general not a Markov-modulated Poisson process any more. Do characteristics of $\out_\ti$, such as the mean or variance evolution, stay invariant under approximate marginal simulation? A full, but often intractable characteristic is the path measure of $\out_\ti$. The ACID provides a partial characteristic that exceeds first- and second-order analysis but is less complex than the path measure. It can assist as a discrimination tool to detect dissimilar approximate filters. We demonstrate this using the Hawkes process.

\section{Method}
Let $\out_\ti$ be a counting process. Its canonical filtration is $\sigmalg_\ti^\out = \sigma(\out_s: s \le \ti)$. Denote by $\CIntensity_\ti$ the $\sigmalg_\ti^\out$-intensity of the process $\out_\ti$. We refer to it as the \textit{conditional intensity} (CI).
\subsection{The backward recurrence time parametrization}
Throughout this work we consider a counting process $\out_\ti$ whose conditional intensity $\CIntensity_\ti$ is parametrized by a process $(\intime(\ti), \state(\ti))$ in the following form. The scalar $\intime$ is the time since the latest jump of $\out$, i.e., the \textit{backward recurrence time} \cite{Cox.1962}.
The second component $\state \in \R^\num$ is a (possibly multi-variate) \textit{sufficient statistic}, that possesses three properties
\begin{enumerate}[leftmargin=*, widest=7bis), align= left]
    \item[(A1)] it is constant between jumps, i.e. $\state(\ti) = \state(\ti - \intime(\ti))$
    \item[(A2)] there is a deterministic $\cintensity$, satisfying $\CIntensity_\ti = \cintensity(\intime(\ti), \state(\ti))$.
    \item[(A3)] there is a deterministic $\target$, satisfying $\target(\intime(\ti-), \state(\ti-)) = \state(\ti)$ at jump times $\ti$ of $\out$.
\end{enumerate}
We further request $(\intime(\ti), \state(\ti))$ to be ergodic in the following. An example of a non-ergodic process $(\intime(\ti), \state(\ti))$ satisfying the three properties was investigated in \cite{VereJones.1984}.
The process $(\intime(\ti), \state(\ti))$, denoted \textit{BReT-P}, combines three conditions which jointly turn it into a piecewise-deterministic Markov process \cite{Davis.1984}:
\begin{enumerate}[leftmargin=*, widest=7bis), align= left]
\item[(B1)] The evolution equation of $(\intime, \state)$ between jumps reads
\begin{equation}
  \dot\intime = 1, \, \dot\state = 0.
  \label{eq:process_equation}
\end{equation}
Thus the process evolves deterministicly from the current state when there are no jumps.
\item[(B2)] Jumps of $(\intime(\ti), \state(\ti))$ and $\out$ occur simultaneously with intensity $\CIntensity_\ti = \cintensity(\intime(\ti), \state(\ti))$, only depending on the current state, not the history.
\item[(B3)] At jumps, the backwards recurrence time is updated to $\intime(\ti) = 0$ and the sufficient statistic targets $ \state(\ti) = \target(\intime(\ti-), \state(\ti-))$. The jump targets only depend on the current state.
\end{enumerate}
Next, we show that the CI of a Markov-modulated Poisson process is of this form. After the joint asymptotic distribution of $(\intime, \state)$ is found (see section \ref{sub:masterEq}), the ACID is determined via the transform $\cintensity(\intime, \state)$ (see section \ref{sub:Numeric_ACID}).
\subsection{Class of Markov-modulated Poisson processes}
\label{sub:Markov-modulated}
Let $\msg$ be a continuous time Markov chain (CTMC) with finitely many states in $\msgStateSpace$ and $\intensityFct: \msgStateSpace \to [0, \infty)$ an intensity mapping.
Let $\out$ 
be a doubly stochastic Poisson process with intensity $\intensity_\ti = \intensity(\msg_\ti)$.

Denote by $\MasterEq$ the generator of $\msg$. Then the filtering distribution $\CProb_\ti(\msgState) := \Prob[\msg_\ti = \msgState\vert \sigmalg^\out_\ti], \msgState \in \msgStateSpace$ evolves as follows, according to Snyder \cite{Snyder.1991},
\begin{equation}
    \frac{\dd}{\dd \ti} \CProb_\ti(x) = (\MasterEq \CProb_\ti)(x) -  (\intensityFct(x)-\CIntensity_\ti) \CProb_\ti(x)
    \label{eq:between_jumps}
\end{equation}
between jumps and is updated to
\begin{equation}
    \CProb_\ti(\msgState) = \frac{\intensityFct(\msgState)\CProb_{\ti-}(\msgState)}{\CIntensity_{\ti-}} 
    \label{eq:at_jumps}
\end{equation} 
if $\out_\ti = \out_{\ti-} + 1$. Taken together, the Snyder filter reads
\begin{equation}
    \dd \CProb_\ti(x) = (\MasterEq \CProb_\ti)(x) \dd \ti + \frac{(\intensityFct(x) - \CIntensity_\ti) \CProb_\ti(x) }{\CIntensity_\ti} \{ \dd \out_\ti - \CIntensity_\ti \dd \ti \}.
    \label{eq:filter}
\end{equation}
By means of the filtering distribution, the conditional intensity is computed as $$\CIntensity_\ti= \E [\intensity(\msg_\ti)\vert \sigmalg^\out_\ti] = \sum_\msgState \intensityFct(\msgState)\CProb_{\ti}(\msgState).$$ It is also denoted as the filter mean or causal conditional mean estimate.

We now consider $\state(\ti) := \CProb_{\ti} \in [0,1]^{\vert \msgStateSpace \vert}$ at jump times and $\state(\ti) = \state(\ti - \intime(\ti))$ else, and derive that the process $(\intime(\ti), \state(\ti))$ is a BReT-P of $\CIntensity_\ti$. In order to do so, we introduce auxiliary functions $\CIprojection, \auxODE$ and $\jtarget$ to define the functions $\cintensity$ and $\target$ in this context. Let $\CIprojection(\cprob):=\sum_{\msgState}\intensityFct(\msgState)\cprob(\msgState)$ be the mean functional. Observe that, by the relation $\CIntensity_\ti = \CIprojection(\CProb_{\ti})$, the ODE system \eqref{eq:between_jumps} is closed and autonomous. Denote by $\intime \mapsto \auxODE(\intime, \cprob)$ its solution  with initialization $\CProb_0 = \cprob \in [0,1]^{\vert \msgStateSpace \vert}$.
The jump update $\jtarget \colon \R^{\vert \msgStateSpace \vert} \to \R^{\vert \msgStateSpace \vert}$ is $\jtarget(\cprob)(x) = \frac{\intensityFct(\msgState)\cprob(\msgState)}{\CIprojection(\cprob)}$. Then $\cintensity := \CIprojection \circ \auxODE$ and $\target := \jtarget \circ \cintensity$ satisfy the properties (A2) and (A3).

\begin{remark}
It is crucial to note that we can indeed consider $\out_\ti$ to be jumping with intensity $\CIntensity_\ti$ instead of $\intensity_\ti$. In condition (B2), if we replaced $\CIntensity_\ti$ by $\intensityFct(\msg_\ti)$ we would lose the self-contained description of $\CIntensity_\ti$ that even disposes of $\out_\ti$ itself. To this end, in process equation \eqref{eq:filter}, replace $\out_\ti$ by $\tilde{\out}_\ti$, where $\tilde{\out}_\ti$ is a self-exciting counting process with intensity $\CIntensity_\ti$. The processes $\tilde{\out}_\ti$ and $\out_\ti$ are equal in distribution. So instead of $\out_\ti$ we consider $\tilde{\out}_\ti$ in the first place and drop the tilde for convenience.

The fact that the Snyder filter is a piecewise-deterministic Markov process has been observed before \cite{Parag.2017b}. For the particle interpretation of this fact we refer to \cite{Parag.2017}.

\end{remark}
\subsection{Sufficient state variables of joint Markovian progression}
\label{sub:MarkovianProgression}
We can abstract the BReT-P in subsection \ref{sub:Markov-modulated} to a more general process class that falls in our considered framework. Consider state variables $(\SuffStat_1(\ti), \dots, \SuffStat_{n_0}(\ti)) = \SuffStat(\ti) \in \R^{n_0}$ that progress as 
\begin{equation}
    \dot\SuffStat(\ti) = \SuffDyn(\SuffStat(\ti))
    \label{eq:SuffODE}
\end{equation}
with the deterministic dynamics $\SuffDyn$ between jumps of $\SuffStat(\ti)$, and at jumps they are updated to
\begin{equation}
    \SuffStat(\ti) = \jtarget(\SuffStat({\ti-}))
    \label{eq:SuffUpdate}
\end{equation}
with the update function $\jtarget = (\jtarget_1, \dots, \jtarget_{n_0}) \colon \R^{n_0} \to \R^{n_0}$.
Let further be $\CIprojection \colon \R^{n_0} \to \R$ a deterministic functional of the state variable, such that $\CIntensity_\ti = \CIprojection(\SuffStat(\ti))$. We call $\SuffStat_1, \dots, \SuffStat_{n_0}$ \textit{sufficient state variables of joint Markovian progression}, because they form a Markov process by arguments analogous to (B1)-(B3) and are sufficient in that the conditional intensity can be computed from them. Denote by $\intime \mapsto \auxODE(\intime, \suffStat^0)$ the solution of eq.~\eqref{eq:SuffODE} with initial value $\SuffStat_0 = \suffStat^0$.  
Suppose that $\jtarget_{n+1} \equiv \suffStat^0_{n+1}, \dots, \jtarget_{n_0}\equiv \suffStat^0_{n_0}$ are constant, i.e., the values $\SuffStat_{n+1}, \dots, \SuffStat_{n_0}$ are reset to the same values $\suffStat^0_{n+1}, \dots, \suffStat^0_{n_0}$ at any jump. We now construct the process $(\intime(\ti), \state(\ti))$. As sufficient statistic it suffices to define $\state(\ti) = (\SuffStat_1(\ti), \dots, \SuffStat_n(\ti))$ at jumps and $\state(\ti) = \state(\ti - \intime(\ti))$ else, automatically satisfying (A1). The extension operator $\extension(\state) := (\state, \suffStat^0_{n+1}, \dots, \suffStat^0_{n_0})$ concatenates the current state with the constants. The truncation operator $\updateProj(\suffStat^0) = (\suffStat^0_1, \dots, \suffStat^0_n)$ projects onto the first $n$ components. The functions $\cintensity(\intime, \state) = \CIprojection \circ \auxODE(\intime, \extension(\state)))$ and $\target(\intime, \state) = \updateProj \circ \jtarget(\auxODE(\intime, \extension(\state)))$ satisfy the properties (A2) and (A3).

\begin{remark}\label{rem:zero-states} We return to section \ref{sub:Markov-modulated} and exclude the state variables with constant reset value. Namely, conservation of probability mass and zero-states reduce the dimension of the sufficient statistic.

\begin{enumerate}[label = (\roman*)]
    \item In eq.~\eqref{eq:filter} the evolution equation of the last $\msgState$ can be replaced by the trivial evolution of $\sum_{\msgState} \CProb_\ti(\msgState)$. The value of this sum is constantly $1$. Hence, the number of state variables $\num_0$ is at most $\vert \msgStateSpace \vert -1$. A reparametrization may further decrease the number. 
    \item Call any $\msgState \in \msgStateSpace$ with $\intensityFct(\msgState) = 0$ a \textit{zero-state}. For zero-states, the reset value of the corresponding $\CProb_\ti(\msgState)$ in eq.~\eqref{eq:at_jumps} is $0$, i.e., $\jtarget$ is constant for these components. Hence, the conditional probabilities for the zero-states need not be tracked in the sufficient statistic $\state$. The dimension of $\state$ can be reduced to
    \begin{equation}
        \num \le \vert \{\msgStateSpace  \colon \intensityFct(\msgState) > 0\} \vert -1.
        \label{eq:zero-states}
    \end{equation}
\end{enumerate}
\end{remark}

\subsection{BReT-P examples}
\label{sub:examples}

\begin{table*}
\centering
\begin{tabular}{c||c |c |c |c|c|c |c |c}
     Model& Type& $\SuffStat(\ti)$ & $\num_0$ & $\num$ & $\CIprojection$ & $\auxODE$ & $\SuffDyn$ & $\jtarget$\\
     \hline
     \hline
     Random telegraph&MM& $\CProb_\ti(1)$ &$1$&$0$& $\gain\cprob_1$ & \checkmark & Eq. \eqref{eq:logistic} & $1$\\
     \hline
     Dark current&MM& $\CIntensity_\ti$ &$1$&$1$& $ \id$ &\checkmark&Eq. \eqref{eq:Leakage}&$\CIntensity^{-1}(\CIntensity - \leakage)(\ONState - \CIntensity) + \CIntensity$ \\
     \hline
     Double On&MM& \makecell{$(\DonsoffFrac_\ti,\condmean_\ti, \CProb_\ti(\msgState_3),$\\ $\dots, \CProb_\ti(\msgState_{\vert \msgStateSpace \vert -1}))$ } & $\vert \msgStateSpace \vert - 1$& $1$& $ \gain \z$ &-&\makecell{Ito from \\Eq. \eqref{eq:between_jumps}}&$(\Donsofffrac, 1, 0, \dots, 0)$ \\
     \hline
     Double On Single Off&MM&$(\DonsoffFrac_\ti,\condmean_\ti)$ &$2$&$1$& $\gain \z$ &-&Eq.\eqref{eq:Donsoff_frac}-\eqref{eq:Donsoff_cintensity}&$(\Donsofffrac,1)$\\
     \hline
     Hawkes&SE& $\CIntensity_\ti$ &$1$&$1$& $\id$ &\checkmark&Eq. \eqref{eq:Hawkes}&$\CIntensity + \HawkesJump$\\
     \hline
     Gamma filter&SE& $(\GammaMean_\ti, \GammaVar_\ti)$ &$2$&$2$& $\gain \gammamean$ &-& Eq. \eqref{eq:Gamma_filter}&$(\gammamean, \Gammavar) + (\Gammavar\gammamean^{-1}, \Gammavar^2\gammamean^{-2})$
\end{tabular}
\label{tbl:models}
\caption{The type indicates whether the model is a Markov-modulated (MM) Poisson process or self-exciting (SE) counting process. The minus ($-$) in the column $\auxODE$ indicates that no analytic solution is available for $\intime \mapsto \auxODE(\intime, \suffStat^0)$.}
\end{table*}
We list examples of parametrizations in order to convey the intuition behind the sufficient state variables of joint Markovian progression. The examples introduced here serve as our case studies in section \ref{sec:cs}, highlighting different aspects of the BReT-P method and usage of ACID. In case studies 1) to 3) ACID is used to evaluate the mutual information rate via \eqref{eq:Liptser}. Examples 4) and 5) show how the analysis of ACID exceeds first and second order analysis, both on the CI level and the level of $\out_\ti$. Contrasting examples 2) and 4) illustrates how ACID can discriminate approximate filters. The table \ref{tbl:models} summarizes the computational details of the examples.
\subsubsection{Random telegraph model}
\label{ex:random_telegraph} The input $\msg$ is a random telegraph model on $\msgStateSpace := \{0,1 \}$ with On and Off rates $\LeakageOn, \LeakageOff$ and $\intensityFct(\msgState) = \gain\msgState$. By remark \ref{rem:zero-states}(i) $\CProb_\ti(1)$ is a sufficient state variable (i.e., $\num_0 = 1$). It evolves according to eq.~\eqref{eq:between_jumps} as
        \begin{equation}
        \dot \CProb_\ti(1) = \LeakageOn - (\LeakageOff + \LeakageOn + \gain) \CProb_\ti(1) + \gain\CProb_\ti(1)^2.
        \label{eq:logistic}
    \end{equation}
    For the conditional intensity, we get 
    $$ \CIntensity_\ti = c \cdot \CProb_\ti(1) + 0\cdot (1-\CProb_\ti(1))$$
    which means $\CIprojection(\cprob_1) = c\cprob_1$. The zero-state reduces the dimension. Indeed, by eq.~\eqref{eq:at_jumps} the reset is $\jtarget(\cprob_1) = 1$, so the dimension of $\state$ can be chosen $\num = 0$ as discussed in subsection \ref{sub:MarkovianProgression}, compare also eq.~\eqref{eq:zero-states}. Consequently, the scalar $\intime(\ti)$ suffices as BReT-P. Let $\intime \mapsto \auxODE(\intime)$ be the solution of the ODE with initial value $\cprob_1 = 1$, then $\CIntensity_\ti = \cintensity(\intime(\ti)) = \CIprojection \circ \auxODE(\intime(\ti)) = \gain \auxODE(\intime(\ti))$.\\
    The BReT-P method helps in analyzing how $\IR(\msg, \out)$ depends on the system parameters $\LeakageOn, \LeakageOff, \gain$, see paragraph \ref{cs:rand_tele}, with the goal to answer questions of optimality under constraints $0 < \LeakageOn \le \OnConstraint, 0 < \LeakageOff \le \OffConstraint$.
\subsubsection{Random telegraph model with dark current}
\label{ex:leakage} Let $\msg$ be a random telegraph model as before, but this time, $\intensityFct(0) =:\leakage > 0 $ introduces a dark current. Define $\ONState := \intensityFct(1) > \intensityFct(0)$ and the amplitude $\dynamicRange := \ONState - \leakage$. By remark \ref{rem:zero-states}(i) again $\num_0 = 1$. Choose $\CIntensity_\ti = \leakage +  \CProb_\ti(1) \dynamicRange$ as sufficient state variable of Markovian progression. Being an affine-linear transform of eq.~\eqref{eq:filter}, it evolves as
        \begin{equation} \begin{split}
        \dd \CIntensity_\ti = \left\{\LeakageOn\dynamicRange - (\LeakageOff + \LeakageOn + \dynamicRange)(\CIntensity_\ti - \leakage) \right.\\
        \left.+ (\CIntensity_\ti - \leakage)^2\right\} \dd \ti + \frac{(\CIntensity_{\ti-} - \leakage)(\ONState-\CIntensity_{\ti-})}{\CIntensity_{\ti-}} \dd \out_\ti.
        \end{split}
        \label{eq:Leakage}
    \end{equation}
     (By their affine-linear relation, $\CIntensity_\ti$ and $\CProb_\ti(1)$ are equivalent choices.)
    From eq.~\eqref{eq:zero-states} we can only conclude $\num \le 1$. Thus, we have $(\intime(\ti), \state(\ti))$ with $\state(\ti) = \CIntensity_\ti$ at jumps as BReT-P.
    BReT-P helps in investigating how dark current alters $\IR(\msg, \out)$, see paragraph \ref{subsec:leakage}.
    
\subsubsection{CTMC with two On states}\label{ex:Donsoff} 
\begin{figure}
    \centering
    \begin{tikzpicture}[->, >=latex', auto, semithick, node distance=1/3*\the\myfigwidth]
\tikzstyle{every state}=[fill=white,draw=black,thick,text=black,scale=1]
\node[state]    (A)                     {Off};
\node[state]    (B)[right of=A]   {On};
\node[state]    (C)[right of=B]   {On};
\path
(A) edge     node{$\DonsoffOn$}     (B)
(B)    edge    node{$\DonsoffR$}      (C)
 (C)   edge[bend left, below]     node{$\DonsoffOff$}     (A);
\end{tikzpicture}
\caption{\textbf{State diagram for the Double On Single Off (DOnSOff) model.} The three-state model is Markovian. A refractory second active state realizes a non-exponential sojourn time in On. For $\DonsoffR = \DonsoffOff$ the sojourn time is an Erlang distribution. The Off sojourn time remains exponential.}
\label{fig:Donsoff_states}
\end{figure}
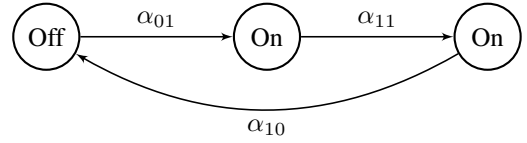

Let $\msg$ be an arbitrary ergodic CTMC with $n$ states and $\intensityFct: \msgStateSpace \to \{ 0,\gain \}$ be binary with two active, i.e., non-zero states $\msgState_1, \msgState_2$. We refer to the model as double On (DOn). Then introduce the conditional probability of being in an active state
    $$\condmean_\ti = \CProb_\ti(\msgState_1) + \CProb_\ti(\msgState_2)$$ and $\DonsoffFrac_\ti:= \CProb_\ti(\msgState_1)/\condmean_\ti$ the contribution of $\msgState_1$ to this conditional probability. Then $\SuffStat_\ti := (\DonsoffFrac_\ti, \condmean_\ti, \CProb_\ti(\msgState_3), \dots, \CProb_\ti(\msgState_{\vert \msgStateSpace \vert - 1}))$ are sufficient variables of joint Markovian progression. The reset value at jumps is $(\DonsoffFrac_{\ti-}, 1, 0, \dots, 0)$. Since all but the first component are set to a constant at jumps, the sufficient statistic $\state(\ti) = \DonsoffFrac_\ti$ at jumps has dimension $\num = 1$. The progression of  $\SuffStat_\ti$ can be found by the chain rule from eq.~\eqref{eq:filter}. We only elaborate on it for the circular CTMC with double On state and single Off state (DOnSOff), depicted in figure \ref{fig:Donsoff_states}.
Let $\DonsoffOn, \DonsoffR, \DonsoffOff$ be the transition rates of going from inactive to first active, first to second active and second active back to inactive, then the generator matrix is of the form
\begin{equation}
    \MasterEq = \begin{pmatrix}
    -\DonsoffR & \DonsoffR& 0\\
    0 & -\DonsoffOff & \DonsoffOff\\
    \DonsoffOn & 0 & -\DonsoffOn
    \end{pmatrix}.
\end{equation}
    The evolution of the conditional probabilities $\CProb_\ti(\msgState_1)$ and $\CProb_\ti(\msgState_2)$ is
    \begin{align}
        &\dot\CProb_\ti(\msgState_1) \nonumber\\ &= \DonsoffOn(1 - \condmean_\ti) - \DonsoffR\CProb_\ti(\msgState_1) - \gain(1 - \condmean_\ti))\CProb_\ti(\msgState_1)\\
        &\dot\CProb_\ti(\msgState_2)\nonumber\\ &= \DonsoffR\CProb_\ti(\msgState_1) - \DonsoffOff\CProb_\ti(\msgState_2) - \gain(1 - \condmean_\ti))\CProb_\ti(\msgState_2).
    \end{align}
    Via the function $\varphi(\cprob_1, \cprob_2) = (\cprob_1/(\cprob_1 + \cprob_2), \cprob_1 + \cprob_2) = (\Donsofffrac, \z)$ this transforms to
    \begin{align}
    \dot \DonsoffFrac &= - \DonsoffR \DonsoffFrac + \DonsoffOn \frac{(1-\condmean)(1-\DonsoffFrac)}{\condmean} + \DonsoffOff \DonsoffFrac(1-\DonsoffFrac) \label{eq:Donsoff_frac}\\
    \dot \condmean&= - \gain (1-\condmean) \condmean + \DonsoffOn (1-\condmean) - \DonsoffOff (1-\DonsoffFrac) \condmean. \label{eq:Donsoff_cintensity}\
\end{align}
    The update function according to eq.~\eqref{eq:at_jumps} is $\jtarget(\Donsofffrac, \z) = (\Donsofffrac, 1)$, i.e., constant in the second component. Figure \ref{fig:Donsoff_traj} visualizes how trajectories for different initial values $[\DonsoffFrac(\intime), \condmean(\intime)]_{\intime = 0} = [\state, 1]$ evolve. The coordinate system shows the plane $(\cprob_1, \cprob_2)$, while the grid indicates the transformed radial-like coordinates $(\Donsofffrac, \z)$.
    \begin{figure}
        \centering
        \input{Fig1_220109.tex}
        \caption{\textbf{The Double On Single Off input.} Trajectories $(\CProb_\intime(\msgState_1), \CProb_\intime(\msgState_2))$ evolve in the $(\cprob_1, \cprob_2)$-plane with $\cprob_1+ \cprob_2 \le 1$. The grey grid indicates the change of coordinates $(\Donsofffrac, \z) = (\cprob_1/(\cprob_1+ \cprob_2), \cprob_1+ \cprob_2)\in [0,1]^2$. The initial value is $[\Donsofffrac_0, \z_0] = [\state, 1]$. Dots indicate the temporal evolution, i.e. trajectory values at equally spaced time points. The equilibrium is marked by the cross. Parameter values were $\DonsoffOn = 0.4, \DonsoffR =\DonsoffOff = 0.8, \gain = 1, \state = 0.2, 0.4, 0.6$}
        \label{fig:Donsoff_traj}
    \end{figure}
    The DOnSOff model is equivalent to a binary semi-Markov process with exponential sojourn time in the Off state, while the sojourn time in the On state is the convolution of two exponential distributions. It serves as an example of a non-Markovian binary input to the Poisson channel whose $\IR(\msg, \out)$ is compared to the Markov case, see case study $\ref{cs:donsoff}$.
    
\subsubsection{Hawkes process}\label{ex:Hawkes} The Hawkes process \cite{HAWKES.1971} is a self-exciting counting process whose conditional intensity evolves like
\begin{equation}
    \hat\intensity_\ti = \HawkesBase + \int_0^\ti \HawkesKernel(s, \ti) \dd \out_s,
    \label{eq:general_Hawkes}
\end{equation}
where $\HawkesBase \geq 0$ is a base intensity. For the standard Hawkes process with exponential kernel $\HawkesKernel(s,\ti) = \HawkesJump e^{-\HawkesDecay(\ti - s)}$, the CI turns out to be of Markovian progression \cite{Oakes.1975, Daley.2003}
\begin{equation}
\dd \hat\intensity_\ti = -\HawkesDecay(\hat\intensity_\ti - \HawkesBase) \dd \ti + \HawkesJump \dd \out_\ti.
\label{eq:Hawkes}
\end{equation}
The sufficient statistic $\state(\ti)$ can be chosen as $\CIntensity_\ti$ at jump times. The functions $\cintensity(\intime, \state) = \HawkesBase + e^{-\HawkesDecay \intime} (\state - \HawkesBase)$ and $\target(\intime, \state) = \cintensity(\intime, \state) + \HawkesJump$ satisfy (A2) and (A3). The equilibrium distribution of $\CIntensity_\ti$, i.e., the ACID, is mentioned in the literature \cite{Oakes.1975,Daley.2003}, partly because it helps in realizing a stationary Hawkes process. A Hawkes process whose initial intensity value is drawn from ACID is in the stationary regime from the start. We demonstrate how ACID contains more information than the mean and variance of $\CIntensity_\infty$, see case study \ref{subsec:Hawkes}. Oakes \cite{Oakes.1975} suggested that the ACID can be found by iteratively solving an integral equation, but does not elaborate more on it. Daley \& Vere-Jones \cite[7.2.5 (iii)]{Daley.2003} provide a manual for setting up the integral equation and solving it with the method of steps. We are not sure whether Oakes meant this method by "iterative solution". We provide a different approach using BReT-P and a fixed point iteration.

Besides being an interesting counting process model on its own that has seen a variety of applications in different fields of research, the Hawkes process can also be regarded as an approximate filter. It is obtained by projection onto the class of linear estimators under the quadratic criterion and we therefore also call it the optimal linear estimator. In case study \ref{cs:Hawkes filter} we elaborate on this and employ the ACID to discriminate between the exact Snyder filter and the optimal linear filter in \ref{cs:Hawkes_comparison}.

\subsubsection{The Gamma filter} \label{ex:Gamma} The Gamma filter \cite{Zechner.2014, Zechner.2016} is an approximate filter obtained from conditional moment closure. It departs from the linear modulation by a Cox-Ingersoll-Ross process \cite{Cox.1985} with stationary CIR-mean $\GammaMu$ and CIR-autocovariance function $\GammaSigma e^{-\GammaDecay \ti}$, i.e., $\msg_\ti \sim \mathrm{CIR}(\GammaMu, \GammaSigma, \GammaDecay)$ and $\out_\ti$ is a doubly stochastic Poisson process with $\intensity_\ti = \gain \msg_\ti$. The Gamma filter is the assumed density filter, where $\msg_\ti\vert \out_{[0,\ti]}$ is assumed to be Gamma distributed. With two degrees of freedom for the Gamma distribution, two equations sufficiently describe the filter. One governs the mean, one governs the variance.
    Expressing the Gamma's third centered moment in terms of mean and variance justifies the replacement $\E[(\msg_\ti - \GammaMu)^3 \vert \sigmalg^\out_\ti] = \frac{2\E[(\msg_\ti - \GammaMu)^2 \vert \sigmalg^\out_\ti]^2}{\E[\msg_\ti \vert \sigmalg^\out_\ti]}$. Consider an approximate marginal simulation of $\out_\ti$ that uses the Gamma filter. This yields a self-exciting process $\out_\ti$ with $(\GammaMean_\ti, \GammaVar_\ti)$ mimicking $(\E[\msg_\ti \vert \sigmalg^\out_\ti], \E[(\msg_\ti - \GammaMu)^2 \vert \sigmalg^\out_\ti])$, the CI $\CIntensity_\ti = \gain \GammaMean_\ti$ and
    \begin{align}
        \dd\GammaMean_\ti &= \{-\GammaDecay (\GammaMean_\ti - \GammaMu) - \gain \GammaVar_\ti \} \dd \ti + \frac{\GammaVar_{\ti-}}{\GammaMean_{\ti-}}\dd \out_\ti \nonumber\\
        \dd\GammaVar_\ti &= \{-2\GammaDecay (\GammaVar_\ti - \frac{\GammaSigma}{\GammaMu}\GammaMean_\ti) - 2\gain\frac{\GammaVar_\ti^2}{\GammaMean_\ti} \} dt + \frac{\GammaVar_{\ti-}^2}{\GammaMean_{\ti-}^2}\dd \out_\ti.
        \label{eq:Gamma_filter}
    \end{align}
    It is instructive to contrast approximate marginal simulation of $\out_\ti$ with estimation of $\msg_\ti$. Both perspectives can make use of the equations \eqref{eq:Gamma_filter}. In estimation the process $\out_\ti$ is self-exciting with intractable CI $\gain \E[\msg_\ti\vert \sigmalg^\out_\ti]$, while in approximate marginal simulation $\out_\ti$ is by definition self-exciting with $\gain \GammaMean_\ti$.
    We proceed with the latter perspective.
While the CI $\CIntensity_\ti$ alone is not of Markovian progression, the joint $\SuffStat_\ti = (\GammaMean_\ti, \GammaVar_\ti)$ is, yielding $\num = \num_0 = 2$.

    With $\state(\ti) = (\GammaMean_\ti, \GammaVar_\ti)$ at jump times, the sufficient statistic can be defined.
    
The Gamma filter and optimal linear filter are obtained from different closure schemes. In case study \ref{cs:Gamma} the ACID is employed to discriminate between approximate marginal simulations with either one. The case study of the Gamma filter indicates the limitations of ACID as a discrimination tool for approximate filters. It, however, informs a decision, in which parameter regime to replace the Gamma by the optimal linear filter when approximately computing the information rate.

The examples illustrate how the concept of CI unifies the two pictures:
\begin{enumerate}[label = (\roman*)]
    \item the conditional mean estimate in a Markov-modulated Poisson process and
    \item the history-dependent intensity in a self-exciting process, e.g. Hawkes process and marginal simulation via approximate filters.
\end{enumerate}
\subsection{The master equation and stationarity condition}\label{sub:masterEq}
We now aim for the asymptotic distribution of $(\intime(\ti), \state(\ti))$.
The master equation for the piecewise-deterministic Markov process $(\intime(\ti), \state(\ti))$, derived from the process equation $\eqref{eq:process_equation}$, reads \cite{Gardiner.2009}
\begin{equation}
\partial_\ti \density(\ti, \intime, \state) = - \partial_{\intime} \density(\ti, \intime, \state) -  \cintensity(\intime, \state) \density(\ti, \intime, \state).
\label{eq:Master_equation}
\end{equation}
At stationarity the left-hand side vanishes, yielding
\begin{equation}
    \partial_{\intime} \density(\intime, \state) = - \cintensity( \intime, \state) \density(\intime, \state).
    \label{eq:stationarity}
\end{equation}
Let $\Density$ solve this equation for $\Density(0, \state) = 1$. Since eq.~\eqref{eq:stationarity} is linear, the true density then satisfies 
\begin{equation}
  \density(\intime, \state) = \density(0, \state)\Density(\intime, \state).
  \label{eq:factorization}
\end{equation}

\subsection{Integral boundary condition and normalization condition}
The true $\density(0,\state) =: \idensity(\state)$ has to satisfy an influx boundary condition. We state two versions of the condition in the form of integral equations.

\begin{theorem}
For any $\stateSet \in \mathcal B(\R^n)$ it holds
\begin{equation}
    \int_\stateSet \idensity(\state) \dd \state = \int_{\target(\intime, \state) \in \stateSet} \cintensity(\intime, \state)\Density(\intime, \state) \idensity(\state) \dd \state \dd \intime.
    \label{eq:version1}
\end{equation}
\label{theo:version1}
\end{theorem}

\begin{proof}
Let $t >h> 0$. The probability $\Prob[\intime(\ti)\in [0,h), \state(\ti) \in \stateSet]$ can be written in two ways (up to order $o(h)$), first
$$ \int_0^h \int_\stateSet \density(\intime, \state, \ti) \dd \state \dd \intime $$
and second, since we know a jump must have occurred in $(\ti-h, \ti]$
\begin{align*}
 &\int_{\ti - h}^\ti \int \density(\text{jump at }h'\vert \state(h') = \state', \intime(h') = \intime') \times\\
 &\qquad \qquad \density(\intime, \state, h') \Id(\target(\intime', \state') \in \stateSet) \dd \state' \dd \intime'\dd h' \\
 &= \int_{\ti - h}^\ti \int\cintensity(\intime', \state') \density(\intime', \state', h') \Id(\target(\intime', \state') \in \stateSet) \dd \state' \dd \intime'\dd h'.
\end{align*} 
Then dividing by $h$ and letting $h \to 0$ gives the equality
\begin{equation*}
    \int_\stateSet \density(0,\state, \ti) \dd \state = \int_{\target(\intime, \state) \in \stateSet} \cintensity(\intime, \state)\density(\intime,\state, \ti) \dd \state \dd \intime.
\end{equation*}
If we drop the $\ti$ because of stationarity and use eq.~\eqref{eq:factorization} on the right-hand side, we get the result.
\end{proof}

The differential version of this for the special case $n = 1$ is
\begin{theorem}
\label{theo:version2}
For each $\stateto, \statefrom$ let $\intime_i(\stateto, \statefrom)$ for $i = 1, \dots, N(\stateto, \statefrom)$ be an enumeration of the solutions to $\target(\intime, \statefrom) = \stateto$. Assume $\partial_\intime \target(\intime_i(\stateto, \statefrom), \statefrom) \neq 0$ for all $i$.
Here, $N(\stateto, \statefrom)$ is the number of such solutions. Then it holds
\begin{equation}
    \idensity(\stateto) = \left.\int \sum_{i = 1}^{N(\stateto,\statefrom)} \frac{\cintensity(\intime, \statefrom)\Density(\intime, \statefrom)}{\vert \frac{\partial}{\partial \intime} \target(\intime, \statefrom) \vert} \idensity(\statefrom) \right\vert_{\intime = \intime_i(\stateto, \statefrom)} \dd \statefrom .
    \label{eq:version2}
\end{equation}
\end{theorem}
\begin{proof}
Split $\{1, \dots,N(\stateto, \statefrom)\}$ into $\posSet$ and $\negSet$ depending on whether $\intime \mapsto \target(\intime, \statefrom) - \stateto$ has a sign change from $-$ to $+$ or from $+$ to $-$ at $\intime_i(\stateto, \statefrom)$. We choose $\stateSet = (-\infty, \state]$ in eq.~\eqref{eq:version1} and take the derivative $\partial_\state$ on both sides. Then on the left-hand side we get $\idensity(\state)$. For the right-hand side define $\integrand(\intime, \state) := \cintensity(\intime, \state)\Density(\intime, \state) \idensity(\state)$. There exists some choice $\posSet = \{j_1, \dots, j_{\num_+(\stateto, \statefrom)} \}$ or $\posSet = \{j_1, \dots, j_{\num_+(\stateto, \statefrom)-1} \}, \infty = j_{\num_+(\stateto, \statefrom)}$ and $\negSet = \{i_1, \dots, i_{\num_+(\stateto, \statefrom)} \}$ or $\negSet = \{i_2, \dots, i_{\num_+(\stateto, \statefrom)}\}, i_1 = 0$ for which the right-hand side can be computed as 
\begin{align*}
    &\partial_\state\int \sum_{k = 1}^{\num_+(\stateto, \statefrom)}\int_{\intime_{i_k}(\stateto, \statefrom)}^{\intime_{j_k}(\stateto, \statefrom)} \integrand(\intime, \statefrom) \dd \intime \dd \statefrom\\
    &=\int \sum_{k = 1}^{\num_+(\stateto, \statefrom)} - \partial_\state \intime_{i_k}(\stateto, \statefrom) \integrand(\intime_{i_k}(\stateto, \statefrom), \statefrom) \\&\qquad \qquad \quad+ \partial_\state \intime_{j_k}(\stateto, \statefrom) \integrand(\intime_{j_k}(\stateto, \statefrom), \statefrom) \dd \statefrom\\
    &=\int \sum_{i \in \posSet \cup \negSet} \sgn\{\partial_\intime \target(\intime_i(\stateto, \statefrom), \statefrom)\} \partial_\state \intime_{i}(\stateto, \statefrom) \\&\qquad \qquad \qquad \qquad \times\integrand(\intime_{i}(\stateto, \statefrom), \statefrom)\dd \statefrom,
\end{align*}
where we used, that the $\partial_\state$-derivatives vanish for the lower and upper limits $\intime_{i_1}(\stateto, \statefrom) = 0$ and $\intime_{j_{\num_+(\stateto, \statefrom)}}(\stateto, \statefrom) = \infty$.
With
\begin{align*}
    1 &= \partial_\stateto\target(\intime_i(\stateto, \statefrom), \statefrom) \\&= \partial_\intime \target(\intime_i(\stateto, \statefrom), \statefrom) \cdot \partial_\stateto\intime_i(\stateto, \statefrom)\\
    &= \vert \partial_\intime \target(\intime_i(\stateto, \statefrom), \statefrom)\vert \cdot \sgn\{\partial_\intime \target(\intime_i(\stateto, \statefrom), \statefrom)\}\partial_\stateto\intime_i(\stateto, \statefrom)
\end{align*}
the result follows.
\end{proof}
\begin{remark}
Note that the integral equation might become singular \cite{Gohberg.1992} if there exist $\intime_0, \stateto_0, \statefrom_0$, such that $\target (\intime_0, \statefrom_0) = \stateto_0$, but $\partial_\intime \target (\intime_0,\statefrom_0) = 0$. This may occur both at $\intime_0$ with sign changes of $\intime \mapsto \target (\intime_0, \statefrom_0) - \stateto_0$ and without sign changes. In the first case, the equation \eqref{eq:version2} ought to be interpreted as having an arbitrary value under the integral for this $(\stateto_0, \statefrom_0)$-pair. In the second case, the value under the integral in eq.~\eqref{eq:version2} is defined for every $(\stateto, \statefrom)$. However, for $\statefrom$ close to $\statefrom_0$ the derivative $\partial_\intime \target(\intime_i(\stateto_0, \statefrom), \statefrom)$ will approach $0$ and hence a singularity will appear nonetheless at $(\stateto_0, \statefrom_0)$. The existence of a solution for this singular integral equation needs to be carefully checked.
\end{remark}
\begin{remark}
Theorem \ref{theo:version2} can be generalized to more than one dimension.
Let $\path_i = (\path_{i,1}, \path_{i,2}) \colon (\low_i,\up_i) \times \stateSupport_i \to [0,\infty) \times \stateSupport, i = 1, \dots, \numPath$ satisfy $\target(\path_i(t, \stateto)) = \stateto$ with each $\path_i$ injective and differentiable and $[0,\infty) \times \stateSupport = \sqcup_{i = 1}^\numPath \path_i((\low_i,\up_i) \times \stateSupport_i)$ up to a set of Lebesgue measure zero. Then it holds that
\begin{align}
    \idensity(\stateto)  = \sum_{i\colon \stateto \in \stateSupport_i } \int_{\low_i}^{\up_i} &\vert \det D\path_i(t, \stateto) \vert \cintensity(\path_i(t, \stateto)) \times \nonumber\\
    &\Density(\path_i(t, \stateto))\idensity(\path_{i,2}(t, \stateto)) \dd t.
\end{align}
\end{remark}

\begin{theorem}
The normalization of $\idensity$ is dictated by the mean intensity, i.e.,
\begin{equation}
 \int \idensity(\state) \dd \state = \E[\CIntensity_\infty].
 \label{eq:normalization_constant}
\end{equation}

\end{theorem}

\begin{proof}
Using $\lim_{\intime \to \infty}\density(\intime, \state) = 0$ and eq.~\eqref{eq:stationarity}, we compute
\begin{align*}
	\int \idensity(\state) \dd  \state &= \int \density(0, \state) - \lim_{\intime \to \infty}\density(\intime, \state) \dd  \state\\
	& = \int \int_0^\infty -\partial_\intime \density(\intime, \state) \dd \intime \dd \state\\
	&= \int\int_0^\infty \cintensity(\intime, \state) \density(\intime, \state) \dd \intime \dd \state
\end{align*}
and recognize the right hand side as $\E[\CIntensity_\infty]$.
\end{proof}

We emphasize that the strong advantage of the BReT-P lies in aligning all probability inflow terms in the master equation at $\intime = 0$. Our method provides the asymptotic distribution of $(\intime, \theta)$. If the sufficient state variable $\SuffStat$ itself is chosen as the domain variable of the asymptotic distribution, then the stationarity equation \eqref{eq:stationarity} grows wider: influx terms will be needed for any values of $\SuffStat$ that jumps can anticipate. The equation will assume a difference-differential form, e.g. \cite[7.2.5 (iii)]{Daley.2003} or \cite[III.C]{Sinzger.2020}. The BReT-P circumvents this difference-differential formulation for which solution techniques, like method of steps, would be needed in general. A direct solution technique for one-dimensional $\SuffStat$ can handle the difference-differential formulation. It uses a fixed point treatment, similar to eq.~\eqref{eq:version2}. We refer the reader to paragraph \ref{sub:direct} and continue with the BReT-P method here. The appealing simplicity of eq.~\eqref{eq:stationarity} - being autonomous linear ODEs - comes at the cost of an integral boundary condition ($\intime = 0$) which is more involved.

A further advantage of the BReT-P lies in dimension reduction: The pair $(\intime, \state)$ uniquely informs all state variables $\SuffStat_{\num +1}, \dots, \SuffStat_{\num_0}$ with constant reset values, i.e. via $(\intime, \state) \mapsto \auxODE(\intime, \extension(\state))$. As discussed, we may consequently dismiss $\SuffStat_{\num +1}, \dots, \SuffStat_{\num_0}$ in the sufficient statistic $\state$. Eq.~\eqref{eq:SuffODE} - \eqref{eq:SuffUpdate} show that the state variables of Markovian progression $\SuffStat_{1}, \dots, \SuffStat_{\num_0}$ also form a piecewise-deterministic Markov process. In case we apply the Markov theory onto $\SuffStat_{1}, \dots, \SuffStat_{\num_0}$ instead of $(\intime, \state)$, can we dismiss the state variables with constant reset value as well? Generally not, because the mapping $(\SuffStat_{1}(\ti), \dots,\SuffStat_{\num_0}(\ti)) \mapsto (\SuffStat_{1}(\ti), \dots,\SuffStat_{\num}(\ti))$ can be non-injective. In that case there exists no unique mapping $(\SuffStat_{1}(\ti), \dots,\SuffStat_{\num}(\ti)) \mapsto(\SuffStat_{1}(\ti), \dots,\SuffStat_{\num_0}(\ti))$. As an example, see fig. \ref{fig:Donsoff_traj} with $\SuffStat_1 = \DonsoffFrac, \SuffStat_2 = \condmean$. The trajectory starting at $\DonsoffFrac(0) = 0.6$ intersects $\Donsofffrac = 0.5$ twice.

\subsection{Discretization}
We discretize the integral boundary condition~\eqref{eq:version1}. To this end, we assume that $\idensity(\state)$ is supported on $\stateSupport \subseteq \R^\num$. We choose a partition $(\statePartition_\ipart)_{\ipart = 1, \dots, \inum}$ with equivalent volume $\vol(\statePartition_\ipart) = \vol(\statePartition_\iPart) =: \weight$. For $\num = 1$, this is an equidistant partition. We discretize $\idensity(\state)$ as
\begin{equation}
 \idensity(\state) = \sum_{\ipart =1}^\inum \idensityCoeff_\ipart \Id_{\statePartition_\ipart}(\state) 
\end{equation}
with unknowns $\idensityCoeff_\ipart$ and discretize $\cintensity(\intime, \state)$, respectively $\Density(\intime, \state)$, as
\begin{equation}
\cintensity(\intime, \state) = \sum_{\ipart =1}^\inum \cintensity(\intime, \state_\ipart) \Id_{\statePartition_\ipart}(\state) , \quad \Density(\intime, \state) = \sum_{\ipart =1}^\inum \Density(\intime, \state_\ipart) \Id_{\statePartition_\ipart}(\state)
\label{eq:discretization}
\end{equation}
for a choice of representatives $\state_\ipart \in \statePartition_\ipart$, e.g. the center of $\statePartition_\ipart$.
Define the border crossing time points $\intime_0(\statefrom) := 0$ and recursively
$\intime_\itime(\statefrom) := \min\{ \intime > \intime_{\itime - 1} : \target(\intime , \statefrom) \in \partial \statePartition_\ipart \text{ for some } \ipart \}$.
Define $\itrap(\itime, \statefrom) :=  \ipart$ if $\target(\intime, \statefrom) \in \statePartition_\ipart$ for $\intime_{\itime-1}(\statefrom) < \intime < \intime_\itime(\statefrom)$. Then for $\stateSet = \statePartition_\ipart$ the equation \eqref{eq:version1} reads
\begin{align}
    \idensityCoeff_\ipart &= \sum_{\iPart = 1}^\inum \left(\sum_{\itime: \itrap(\itime,\statefrom_\iPart) =\ipart }\int_{\intime_{\itime-1}(\statefrom_\iPart)}^{\intime_{\itime}(\statefrom_\iPart)} \cintensity(\intime, \statefrom_\iPart) \Density(\intime, \statefrom_\iPart) \dd \intime \right) \idensityCoeff_\iPart \nonumber\\
    &= \sum_{\iPart = 1}^\inum \left(\sum_{\itime: \itrap(\itime,\statefrom_\iPart) =\ipart } \Density(\intime_{\itime-1}(\statefrom_\iPart), \statefrom_\iPart ) - \Density(\intime_{\itime}(\statefrom_\iPart), \statefrom_\iPart ) \right) \idensityCoeff_\iPart.
    \label{eq:matrix_version1}
\end{align}
Defining the bracket term as $\OpI_{\ipart, \iPart}$, the equation can be written $\vec\idensityCoeff = \OpI \vec\idensityCoeff$ in matrix form. Observe that by a telescope sum argument and $\Density(0,\statefrom_\iPart) = 1$ the matrix $\OpI$ is left stochastic. If $\OpI$ is quasi-positive, the fixed point equation has a unique non-zero solution by the Perror-Frobenius theorem. Then $\vec\idensityCoeff$ can be approximated by taking a column of a large enough power $\OpI^{2^L}$ and multiplying by the product of weight $\weight$ and normalization constant $\E[\CIntensity_\infty]$.

\subsection{The mutual information rate}
We assume that $\out_\ti$ is Markov-modulated with intensity $\intensity_\ti$ and CI $\CIntensity_\ti$ given by a BReT-P, e.g., as in section \ref{sub:Markov-modulated} or \ref{sub:MarkovianProgression}. Let $\intensity_\ti \to \intensity_\infty$ in distribution. The mutual information rate is
\begin{equation}
    \IR(\msg, \out) = \E[\phi(\intensity_\infty)] - \int \int_0^\infty \phi(\cintensity(\intime, \state)) \density(\intime, \state) \dd \intime \dd \state.
    \label{eq:IR}
\end{equation}
The first term is a finite sum. The outer integral can be approximated by
\begin{equation}
\sum_{\ipart = 1}^\inum \idensity(\state_\ipart) \weight \int_0^\infty  \phi(\cintensity(\intime, \state_\ipart)) \Density(\intime, \state_\ipart) \dd \intime.
\label{eq:IR_discretization}
\end{equation}
Define the partial integral $\cummIR(\ti, \statefrom)$ by
\begin{equation}
    \cummIR(\ti, \statefrom) := \int_0^\ti \phi(\cintensity(\intime, \statefrom)) \Density(\intime, \statefrom) \dd \intime.
\end{equation}
In order to proceed from there, we assume that the BReT-P for $\CIntensity_\ti$ is as described in subsection \ref{sub:MarkovianProgression}.
Then for each $\ipart$, the integral in eq.~\eqref{eq:IR_discretization} is efficiently solved by the joint ODE system, the dot denoting the $\intime$-derivative,
\begin{align}
    \dot \Density(\intime, \state_\ipart) &= - \CIprojection(\suffStat(\intime)) \Density(\intime, \state_\ipart)\label{eq:Density}\\
    \dot \suffStat(\intime) &= \SuffDyn(\suffStat(\intime))\\
    \dot \cummIR(\intime, \state_\ipart) &= \phi(\CIprojection(\suffStat(\intime)))\Density(\intime, \state_\ipart).\label{eq:cummIR}
\end{align}
The initial conditions are $\Density(0,\state_\ipart) = 1$, $ \suffStat(0) = (\state_\ipart, \suffStat_{\num + 1}^0, \dots, \suffStat_{\num_0}^0)$ and $\cummIR(0,\state_\ipart) = 0$.

\begin{remark}[$\cummIR$ converges to the information rate exponentially fast]\label{rem:MI_convergence}
If $\cintensity(\intime, \state) \geq \cintensity_0 > 0$ uniformly over $\intime, \state$, then $\Density(\intime, \state) \le e^{-\cintensity_0\intime}$. Let furthermore $\cintensity(\intime, \state) \le R(\state)$. Then
\begin{align*}
    \vert \cummIR(\infty, \statefrom) - \cummIR(\ti, \statefrom) \vert &\le \int_\ti^\infty \vert \phi(\cintensity(\intime, \statefrom)) \vert e^{-\cintensity_0 \intime} \dd \intime\\
    &= \frac{\max\{\phi(R(\statefrom)), \frac{1}{e}\}}{\cintensity_0} e^{-\cintensity_0 \ti}.
\end{align*}
In order to obtain exponential convergence of $\int \idensity(\statefrom) \cummIR(\ti, \statefrom) \dd \statefrom$ to the integral in eq.~\eqref{eq:IR} for $\ti \to \infty$, it needs to holds that
$$ \int \idensity(\statefrom)\max\{\phi(R(\statefrom)) , \frac{1}{e}\} \dd \statefrom < \infty.$$
Sufficient conditions are, for instance, if $R$ is bounded or if $\state \mapsto R(\state)$ is linear in $\state$ and the second moment of $\idensity$ exists.
\end{remark}
As a summary, a numerical approximation of the mutual information rate can be obtained by solving for $\cummIR(\infty, \state_\ipart)$ via eq.~\eqref{eq:Density} - \eqref{eq:cummIR} and $\idensity(\state_\ipart)$ via eq.~\eqref{eq:matrix_version1}. The weights in eq.~\eqref{eq:matrix_version1} can be determined on the fly while solving eq.~\eqref{eq:Density} - \eqref{eq:cummIR}.
\subsection{Numerical approximation of the ACID}
\label{sub:Numeric_ACID}
The ACID is obtained as the distribution $\acid$ of $\cintensity(\intime, \state)$, where $(\intime, \state)$ is distributed according to the density $\density(\intime, \state)$.
We discretize $(0,\infty)$ with mesh size $\Delta \cintensity$ and compute the weights
\begin{align*}
    \mdensityCoeff_\ipart &:= \Prob[\mintervalBound_{\ipart - 1} \le \cintensity(\intime, \state) \le \mintervalBound_{\ipart}]\\ &= \int \int_0^\infty \Id_{(\mintervalBound_{\ipart - 1}, \mintervalBound_{\ipart}]}(\cintensity(\intime, \state)) \Density(\intime, \state) \idensity(\state) \dd \intime \dd \state.
\end{align*}
If we define $\intime^{(\cintensity)}_\itime(\statefrom)$ similarly as $\intime_\itime(\statefrom)$, namely by $\intime^{(\cintensity)}_0(\statefrom) = 0$ and recursively $\intime^{(\cintensity)}_\itime(\statefrom) := \min\{ \intime > \intime^{(\cintensity)}_{\itime - 1}(\statefrom): \cintensity(\intime, \statefrom) = \mintervalBound_\ipart \text{ for some } \ipart \}$, then
\begin{equation}
   \mdensityCoeff_\ipart \approx \weight \cdot \sum_{\iPart = 1}^\inum \idensity(\statefrom_\iPart) \sum_{\itime: \iTrap(\itime,\statefrom_\iPart) =\ipart } \int_{\intime^{(\cintensity)}_{\itime-1}(\statefrom_\iPart)}^{\intime^{(\cintensity)}_{\itime}(\statefrom_\iPart)} \Density(\intime, \statefrom_\iPart) \dd \intime,
    \label{eq:mdensityCoeffAppro}
\end{equation}
where $\iTrap(\itime, \statefrom) :=  \ipart$ if $\target(\intime, \statefrom) \in \statePartition_\ipart$ for $\intime_{\itime-1}(\statefrom) < \intime < \intime_\itime(\statefrom)$.
We further have
\begin{equation}
    \Delta \cintensity \cdot \acid(\aci) \approx \sum_{\ipart = 1}^\infty \mdensityCoeff_\ipart\Id_{(\mintervalBound_{\ipart - 1}, \mintervalBound_{\ipart }]}(\aci).
\end{equation}

\section{Case studies\label{sec:cs}}
\subsection{\label{cs:rand_tele}Random telegraph permits an analytic solution}
Continuing example \ref{ex:random_telegraph}, we consider the normalized parameters $\RTON := \RTOn/\gain, \RTOFF := \RTOff/\gain$ and, for convenience, drop the tilde again. For simplicity, $\cprob := \cprob_1$. Let $\RTRoot_1< \RTRoot_2$ be the roots of the quadratic equation at equilibrium
\begin{equation}
        0 = \gain(\RTOn - (\RTOff + \RTOn + 1) \RTRoot + \RTRoot^2)
        \label{eq:logistic_equilibrium}.
    \end{equation}
Compute the solution of eq.~\eqref{eq:logistic} that yields
\begin{equation}
   \cprob(\intime) =  \RTRoot_2 - \frac{\Delta \RTRoot}{1 + \frac{1 - \RTRoot_1}{ \RTRoot_2 - 1} e^{-\gain\Delta \RTRoot \intime}}. 
   \label{eq:rt:cintensity}
\end{equation}
 Then the unnormalized density $\Density(\intime)$ is obtained from solving eq.~\eqref{eq:stationarity}
\begin{equation}
    \Density(\intime) =  e^{-\gain \RTRoot_1 \intime}\frac{\RTRoot_2 - 1}{\Delta\RTRoot} + e^{-\gain \RTRoot_2 \intime}\frac{1 - \RTRoot_1}{\Delta\RTRoot}
    \label{eq:mixture_exponentials}
\end{equation}
and $\idensity = \E[\intensity_\infty] = \frac{\gain \RTOn}{\RTOn + \RTOff}$.
Using $\phi(\gain x) = \gain \phi(x) + x \phi(\gain)$ the mutual information rate can be written as
\begin{align}
   \IR (\msg, \out) =  &- \gain \idensity  \int_0^\infty \phi(\cprob(\intime)) \Density(\intime) \dd \intime \label{eq:reparametrization}\\
   = &- \frac{\gain\RTOn}{\RTOn + \RTOff} \int_0^\infty \phi\left(\RTRoot_2 - \frac{\Delta \RTRoot}{1 + \frac{1 - \RTRoot_1}{ \RTRoot_2 - 1} e^{-\Delta \RTRoot \intime}}\right) \times \nonumber\\
   &\times \left[e^{- \RTRoot_1 \intime}\frac{\RTRoot_2 - 1}{\Delta\RTRoot} + e^{- \RTRoot_2 \intime}\frac{1 - \RTRoot_1}{\Delta\RTRoot} \right] \dd \intime.
\end{align}
The linear time scaling $\intime \mapsto \gain \intime$ was used in the second equality.
We identify eq.~\eqref{eq:reparametrization} as a reparametrization of the integral representation previously reported \cite{Sinzger.2020} for $\RTRoot = \RTRoot_1$ and $\CProb_\ti(1) \overset{\mathrm{d.}}{\to} \condmean$
 \begin{equation}
   \IR (\msg, \out) = -\E[\gain\phi(Z)] = -\gain \int_{\RTRoot}^1 \phi(z) \pi_Z(z) \dd z.
   \label{eq:integral_ISIT}
 \end{equation}
The link can be obtained via the transformation rule applied on the transformation $\cprob\colon [0,\infty)\to (\RTRoot, 1]$, see Appendix~\ref{app:repara}.
The reparametrization~\eqref{eq:reparametrization} has the advantage that vertical asymptotes of the integrand at $z = \RTRoot$ can be avoided. Furthermore, the integral bounds do not depend on the system parameters $\RTOn, \RTOff, \gain$, allowing more uniformly chosen integral bounds in the numerical approximation, compare remark \ref{rem:MI_convergence}.
\subsubsection{Computation of partial derivatives}
Another advantage lies in enabling the computation of partial derivatives. Let us pick up the tilde again. In order to answer questions of optimality of $\IR(\msg, \out)$, the partial derivatives of $\IR(\msg, \out) = \IR(\RTON, \RTOFF, \gain)$ are relevant. Note the relation $\partial_1 \IR(\RTON, \RTOFF, \gain) := \partial_{\RTON} \IR(\msg, \out) = \gain \partial_{\RTOn} \IR(\msg, \out)$. In particular the nullclines $[\partial_1 \IR(\RTON, \RTOFF, \gain) = 0]$ and $[\partial_2 \IR(\RTON, \RTOFF, \gain) = 0]$ contain information about optimal points. By the implicit function theorem, the $\RTON$-nullcline $\RTON \mapsto \graphNullcline(\RTON)$ satisfies the ODE
\begin{equation}
\graphNullcline_\gain'(\RTON) = -\frac{\partial_{11}\IR(\RTON, \graphNullcline(\RTON), \gain)}{\partial_{12}\IR(\RTON, \graphNullcline(\RTON), \gain)}.
    \label{eq:IFT}
\end{equation}
Furthermore, to decide on convexity of the $\RTOn$-nullcline, positivity of \begin{equation}
 \graphNullcline_\gain''(\RTON) = \left[ \frac{2 \partial_{11}\IR \partial_{112}\IR}{\partial_{12}\IR^2} -\frac{\partial_{111}\IR}{\partial_{12}\IR} - \frac{\partial_{11}\IR^2 \partial_{122}\IR}{\partial_{12}\IR^3}\right](\RTON, \graphNullcline_\gain(\RTON), \gain)   
 \label{eq:convexity}
\end{equation}
must be checked.
This motivates to compute partial derivatives up to the third order.
The numerical method is exemplified for $\partial_1 \IR(\RTON, \RTOFF, \gain)$. In order to appreciate the reparametrization, we observe that the Leibniz rule for differentiation of the parameter integral in eq.~\eqref{eq:integral_ISIT} fails when there is an asymptote at $\RTRoot_1$. Then the lower boundary term evaluates to $-\infty$. Thus, let us exploit the reparametrization.
Define $\cprob_1(\intime) := \partial_{\RTON} \cprob(\intime)$ and $\density_1(\intime) := \partial_{\RTON} \density(\intime)$ as well as 
\begin{align*}
    \cummIR_1(\intime) := \partial_{\RTON} \cummIR(\intime) &= \partial_{\RTON}  \int_0^\intime - \gain \phi(\cprob(\ti)) \density(\ti) \dd \ti\\
    &=\partial_{\RTON}  \int_0^{\gain \intime} - \phi(\cprob(\ti/\gain)) \density(\ti/\gain) \dd \ti.
\end{align*}
For the evolution of $\cprob_1(\intime)$ one takes advantage of $\dot \cprob_1(\intime) = \partial_1\dot \cprob(\intime)$. The joint evolution of $\density, \cprob, \density_1, \cprob_1, \cummIR_1$ and initial values are given in the Appendix, \ref{app:derivative ODE system}.
 The saturation value $\lim_{\intime \to \infty}\cummIR_1(\intime)$ is the partial derivative $\partial_1 \IR(\RTON, \RTOFF, \gain)$.
\subsubsection{Mutual information in the phase plane}
\begin{figure}
\centering
\input{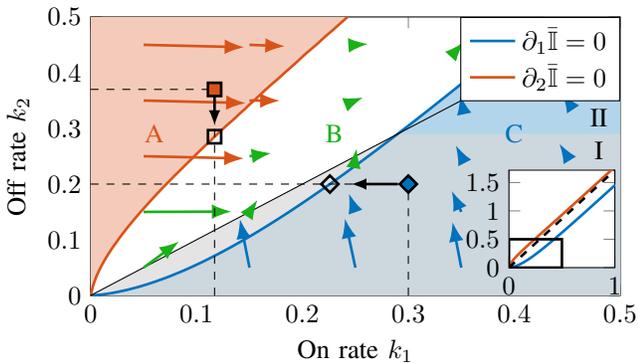}
\caption{\textbf{Phase plane analysis of the mutual information} for the Poisson channel with random telegraph input. Nullclines $\partial_1 \bar{\I} = 0$ and $\partial_2 \bar{\I} = 0$ were evaluated, using $T = 1000$. Colored arrows indicate the gradient of the information rate, calculated alike. Optimization domains are rectangular. Depending on the location of the domain's upper right corner $(r_1, r_2)$, the optimum is assumed on the nullclines $\partial_1 \bar{\I} = 0$ [$(r_1, r_2)$ in C] and $\partial_2 \bar{\I} = 0$ [$(r_1, r_2)$ in A], respectively, or in the interior of region B [$(r_1, r_2)$ in B]. Regions I and II contain all $(r_1, r_2)$, whose optima favor the On and Off state, respectively. In the inset the dashed line $\RTOff = (e-1)\RTOn$ separates the nullclines. The gain was fixed at $\gain = 1$.}
\label{fig:RT_PhasePlane}
\end{figure}

For the analysis we set $\gain = 1$, which is justified by the scaling behavior $\IR(\RTOn, \RTOff, \gain) = \gain \IR(\RTON, \RTOFF, 1) =: \gain \IR(\RTON, \RTOFF)$. For convenience we drop the tilde.
Along the line $(\RTOn, (e - 1)\RTOn)$, both partial derivatives are numerically found not to switch sign. From this we conclude that the nullclines do not intersect. This excludes local optima of $\IR(\RTOn, \RTOff)$. For a rectangular constraint $0 < \RTOn \le \OnConstraint, 0 < \RTOff \le \OffConstraint$, the maximizing pair $(\MaxOn, \MaxOff)$ is consequently always located on one boundary $\RTOn = \OnConstraint$ or $\RTOff = \OffConstraint$. The $(\RTOn, \RTOff)$-plane is split into regions $A, B$ and $C$ by the nullclines. The pair $[\sgn(\partial_1\I), \sgn(\partial_2\I)]$ characterizes the regions: $[1,-1],[1,1],[-1,1]$ on A,B,C. Depending on the location of $(\OnConstraint, \OffConstraint)$, the maximum $(\RTOn^\ast, \RTOff^\ast)$ shows different behavior. If located in A, it holds $\RTOn^\ast = \OnConstraint, \RTOff^\ast < \OffConstraint$, while a location in B enforces $\RTOn^\ast = \OnConstraint, \RTOff^\ast = \OffConstraint$, and finally, $\RTOn^\ast < \OnConstraint, \RTOff^\ast = \OffConstraint$ for $(\OnConstraint, \OffConstraint)$ in region C. As a summary, the maximum is always located in region B or its boundary. Numerically, we find that the $\RTOn$-nullcline crosses the bisection line $(\RTOn, \RTOn)$ at $\intersection$. An evaluation of $\partial_1 \IR(\RTOn, \RTOff)$ for $\RTOn \in (0, \intersection)$ indicates no further intersections. The derivative \eqref{eq:IFT} at $\RTOn = \intersection$ is found to be larger than one. Furthermore, eq.~\eqref{eq:convexity} evaluates to positive values in the region $\{(\RTOn, \RTOff) \in [0, 0.3]^2: \RTOn - 0.07 \le\RTOff \le \RTOn + 0.01\}$. Hence, the isoclines that transit the region are convex, in particular the $\RTOn$-nullcline. The constraint $0 < \RTOff \le \OffConstraint$ for $\OffConstraint \in (0, \intersection)$ then returns a maximum $(\MaxOn, \MaxOff)$ with $\MaxOn > \MaxOff$. This shows that a bandwidth-like constraint can impose an On-favoring maximum complementing the classical result by Kabanov. The sequence of input processes for the Poisson channel that exhaust its capacity is a random telegraph process with $\RTOn, \RTOff \to \infty$ and $\RTOff/ \RTOn \to e-1$. The stationary input distribution favors the Off state, occupying it $1-1/e$ of the time.
\subsubsection{Mutual information rate increases with channel gain}
To answer the question whether $\partial_\gain\IR(\msg, \out)$ is positive, we first observe, see Appendix~\ref{app:partial3},
\begin{equation}
   \partial_3\IR(\RTOn, \RTOff, \gain) =\partial_3\IR\left(\frac{\RTOn}{\gain}, \frac{\RTOff}{\gain}, 1\right).
   \label{eq:scaling}
\end{equation}
Consequently, the analysis of $\partial_3\IR\left(\RTOn, \RTOff, 1\right)$ suffices.
\begin{align}
 &\partial_3\IR\left(\RTOn, \RTOff, \gain\right)\vert_{\gain = 1} \nonumber \\ &= -\partial_\gain\left(\gain \E[\phi(\condmean)]\right)\vert_{\gain = 1}\nonumber\\
 &= \IR\left(\RTOn, \RTOff, 1\right) -\partial_\gain\E[\phi(\condmean)]\vert_{\gain = 1}\nonumber\\
 &= \IR\left(\RTOn, \RTOff, 1\right) - \partial_\gain \int_0^\infty \phi(\cprob(\ti)) \density(\ti) \dd \ti\vert_{\gain = 1}.
 \label{eq:affirm}
\end{align}
The second term is evaluated analogously to ODE \eqref{eq:rho_evolution} - \eqref{eq:Jprime_evolution}.
The numerical evaluation shows that the derivative with respect to the channel gain is non-negative.
We affirm a result in \cite[Corollary 7]{Guo.2008} using eq.~\eqref{eq:affirm}: For $\RTOn, \RTOff \to \infty$ with $\RTOn/\RTOff$ constant, $\condmean$ converges in distribution to a delta distribution at $\frac{\RTOn}{\RTOn + \RTOff}$ independent of $\gain$ and the second term of eq.~\eqref{eq:affirm} vanishes. Consequently, we derive, using in the second equality eq.~\eqref{eq:scaling} and \eqref{eq:affirm},
\begin{align*}
    \partial_3\IR(\RTOn, \RTOff, \gain)\vert_{\gain = 0} &= \lim_{\gain \to 0} \partial_3\IR(\frac{\RTOn}{\gain}, \frac{\RTOff}{\gain}, 1) \\
    &= \lim_{\RTOff \to \infty}\IR(\frac{\RTOn}{\RTOff}\RTOff, \RTOff, 1)\\
    &= \E[\phi(\msg_\infty)] -\phi(\frac{\RTOn}{\RTOn+\RTOff})\\
    &= \E[\phi(\msg_\infty)] - \phi(\E[\msg_\infty]).
\end{align*}

\subsection{Random telegraph with dark current} \label{subsec:leakage}
\begin{figure}
\centering
\input{Fig4_211221.tex}
\caption{\textbf{Random telegraph input with dark current.} ACIDs obtained from Monte Carlo samples and simulation-free computation agree. The density has a vertical asymptote at $\RTRoot_1$ and is non-smooth at $\jtarget_\infty = \jtarget(\RTRoot_1)$. Parameters were $\LeakageOn =  \LeakageOff = 0.1, \leakage = 0.1, \dynamicRange = 1$. Sample size was $10^6$ for the Monte Carlo simulation. Grid point number was $2000$ for the $\idensity$-support $(\jtarget_\infty, \ONState]$ and number of fixed point iterations was $\niter = 12$.}
\label{fig:Leakage_MonteCarlo}
\end{figure}
In the previous example, the dimension $\num$ of the sufficient statistic $\state$ was $0$ and no integral boundary condition was needed. We will proceed with a non-trivial $\state$ when continuing example \ref{ex:leakage}. Similar as in \ref{cs:rand_tele}, define $\tilde{\RTRoot}_1 < \tilde{\RTRoot}_2 $ as the roots of the quadratic equilibrium equation
\begin{equation}
        0 = \RTOn \dynamicRange  - (\RTOff + \RTOn + \dynamicRange) \RTRoot + \RTRoot^2
        \label{eq:leakage:logistic_equilibrium}
    \end{equation}
    and $\RTRoot_i :=\tilde{\RTRoot}_i + \leakage, i = 1,2 $. Obtain
\begin{equation}
   \cintensity(\intime, \state) =  \RTRoot_2 - \frac{\Delta \RTRoot}{1 + \frac{\state - \RTRoot_1}{ \RTRoot_2 - \state} e^{-\Delta \RTRoot \intime}} 
   \label{eq:rt:cintensity}
\end{equation} and
\begin{equation}
    \Density(\intime, \state) =  e^{- \RTRoot_1 \intime}\frac{\RTRoot_2 - \state}{\Delta\RTRoot} + e^{- \RTRoot_2 \intime}\frac{\state - \RTRoot_1}{\Delta\RTRoot}.
    \label{eq:RT_survival_time}
\end{equation}
The number of solutions of $\target(\intime, \statefrom) = \stateto$ is $N(\stateto, \statefrom) \in \{0,1\}$ with
\begin{equation}
    \intime(\stateto, \statefrom) = \Delta \RTRoot^{-1}\left\{\ln\left(\frac{\RTRoot_2 - \jtarget^{-1}(\stateto)}{\jtarget^{-1}(\stateto) - \RTRoot_1}\right) + \ln\left(\frac{\statefrom - \RTRoot_1}{\RTRoot_2 - \statefrom}\right)\right\}.
\end{equation}
Observe that for every $\state$ the trajectory $\cintensity(\cdot, \state)$ is decreasing and $\lim_{\intime \to \infty} \cintensity(\intime, \state) = \RTRoot_1$. Consequently, for any $\intime, \state$
\begin{equation}
    \target(\intime, \state) > \jtarget(\RTRoot_1) = \frac{(\RTRoot_1 - \leakage)(\ONState - \RTRoot_1)}{\RTRoot_1} =: \jtarget_\infty.
\end{equation} We partition the $\idensity(\state)$-support $\stateSupport = (\jtarget_\infty, \ONState]$ into equidistant intervals $(\intervalBound_{\ipart-1}, \intervalBound_{\ipart}]$ with $\intervalBound_{\ipart} = \jtarget_\infty + \ipart \cdot \frac{1 - \jtarget_\infty}{\inum}, \ipart = 1, \dots, \inum$. Choose representatives $\state_\ipart = \frac{\intervalBound_{\ipart} + \intervalBound_{\ipart-1}}{2}$. Then the matrix entries in \eqref{eq:matrix_version1} are given by
\begin{equation}
    \OpI_{\ipart, \iPart} = \Density(\intime(\intervalBound_{\ipart - 1}, \state_\iPart)\vee 0, \state_\iPart) - \Density(\intime(\intervalBound_{\ipart} , \state_\iPart) \vee 0, \state_\iPart)
\end{equation}
with
\begin{equation}
    \Density(\intime(\stateto, \statefrom), \statefrom) = \left( \frac{\jtarget^{-1}(\stateto) - \RTRoot_1}{\statefrom - \RTRoot_1} \right)^{\frac{\RTRoot_1}{\Delta \RTRoot}}\left( \frac{\RTRoot_2 - \statefrom}{\RTRoot_2 - \jtarget^{-1}(\stateto)} \right)^{\frac{\RTRoot_2}{\Delta \RTRoot}}.
\end{equation}
The initial condition $\idensity(\state)$ was found by fixed-point iteration $\OpI^{2^\niter}$ of \eqref{eq:matrix_version1} with $\niter$ iterations. The mutual information rate was computed via \eqref{eq:IR} and \eqref{eq:IR_discretization}.
For \eqref{eq:mdensityCoeffAppro} we compute
\begin{equation}
    \intime^{(\cintensity)}(\stateto, \statefrom) = \Delta \RTRoot^{-1} \left\{ \ln\left(\frac{\RTRoot_2 - \stateto}{\stateto - \RTRoot_1}\right) - \ln\left(\frac{\RTRoot_2 - \statefrom}{\statefrom - \RTRoot_1}\right) \right\}
\end{equation}
and explicitly obtain in \eqref{eq:mdensityCoeffAppro}
\begin{align}
    &\int_{\intime^{(\cintensity)}(\mintervalBound_{\ipart}, \statefrom)}^{\intime^{(\cintensity)}(\mintervalBound_{\ipart-1}, \statefrom)} \Density(\intime, \statefrom) \dd \intime =  \frac{(\RTRoot_2 - \statefrom)^{\frac{\RTRoot_2}{\Delta \RTRoot}}}{ \Delta \RTRoot(\statefrom - \RTRoot_1)^{\frac{\RTRoot_1}{\Delta \RTRoot}}}\times\nonumber\\
    &\qquad \qquad \left[ \frac{(\cintensity - \RTRoot_1)^\frac{\RTRoot_1}{\Delta \RTRoot}}{\RTRoot_1(\RTRoot_2 - \cintensity)^\frac{\RTRoot_1}{\Delta \RTRoot}} + \frac{(\cintensity - \RTRoot_1)^\frac{\RTRoot_2}{\Delta \RTRoot}}{\RTRoot_2(\RTRoot_2 - \cintensity)^\frac{\RTRoot_2}{\Delta \RTRoot}}\right]_{\cintensity = \mintervalBound_{\ipart - 1}}^{\cintensity = \mintervalBound_{\ipart}} .
\end{align}
Fig.~\ref{fig:Leakage_MonteCarlo} shows the agreement of the simulation-free computation with the Monte-Carlo sampling. The ACID has an asymptote at $\RTRoot_1$ and is not differentiable at $\jtarget_\infty = \jtarget(\RTRoot_1)$.

\begin{figure}
    \centering
%
%
\begin{tikzpicture}

\begin{axis}[%
width=0.8*\the\myfigwidth,
height=0.4*\the\myfigwidth,
scale only axis,
xmin=0,
xmax=1,
xlabel style={font=\color{white!15!black}},
xlabel={dark current $\leakage$},
ymin=0,
ymax=0.367879441171442,
ylabel style={font=\color{white!15!black}},
ylabel={information rate},
axis background/.style={fill=white},
legend style={legend cell align=left, align=left, draw=white!15!black}
]
\addplot [color=black, line width = 1pt]
  table[row sep=crcr]{%
0	0.18323460571792\\
0.05	0.151992167080895\\
0.1	0.138022096059258\\
0.15	0.12750940791997\\
0.2	0.119205460789986\\
0.25	0.112225769490153\\
0.3	0.106219961752861\\
0.35	0.100985811575308\\
0.4	0.0963599626981662\\
0.45	0.0922133352082921\\
0.5	0.0884503294952705\\
0.55	0.0850054485521947\\
0.6	0.0818715528011957\\
0.65	0.0790001522607486\\
0.7	0.0763521279622523\\
0.75	0.0738865788498282\\
0.8	0.0715903938450171\\
0.85	0.0694470328287193\\
0.9	0.0674211401655158\\
0.95	0.0655141106138095\\
1	0.0637280097153081\\
};
\addlegendentry{$ \LeakageOn = 0.1, \LeakageOff = 0.1$}

\addplot [color=black, dashed, line width = 1pt]
  table[row sep=crcr]{%
0	0.134369865101935\\
0.05	0.112585178012142\\
0.1	0.100792508607746\\
0.15	0.0921660459639686\\
0.2	0.0852625223832371\\
0.25	0.0795503129759165\\
0.3	0.0746538554355194\\
0.35	0.0704042745983415\\
0.4	0.0666664246317996\\
0.45	0.0633457170690954\\
0.5	0.0603681966417806\\
0.55	0.0576711673982695\\
0.6	0.0551910162193947\\
0.65	0.052916356176528\\
0.7	0.0508666984740027\\
0.75	0.0489798001161725\\
0.8	0.0472370498079893\\
0.85	0.0456184995943172\\
0.95	0.0426973804173094\\
1	0.0413773308789043\\
};
\addlegendentry{$ \LeakageOn = 0.5, \LeakageOff = 0.1$}

\addplot [color=gray, line width = 1pt]
  table[row sep=crcr]{%
0	0.19656577154915\\
0.05	0.145033537541704\\
0.1	0.126389391257974\\
0.15	0.114113245370514\\
0.2	0.10389717470825\\
0.25	0.0956881370714693\\
0.3	0.0888679638589027\\
0.35	0.0830830584761839\\
0.4	0.0780699112017256\\
0.45	0.0736754663918662\\
0.5	0.0697861299643785\\
0.55	0.066296051566018\\
0.6	0.0631611255532982\\
0.65	0.0603467976495562\\
0.7	0.0577803372859631\\
0.75	0.0554314415827712\\
0.8	0.0532815701944025\\
0.85	0.0513033963887699\\
0.9	0.0494778292853715\\
1	0.0461815663747172\\
};
\addlegendentry{$ \LeakageOn = 0.1, \LeakageOff = 0.5$}

\addplot [color=gray, dashed, line width = 1pt]
  table[row sep=crcr]{%
0	0.2956357750538\\
0.05	0.237981154403165\\
0.1	0.209137982033265\\
0.15	0.188578058373118\\
0.2	0.172603169906052\\
0.25	0.159648454936284\\
0.3	0.148731177455383\\
0.35	0.139333796104632\\
0.4	0.1313271713495\\
0.45	0.124304720675642\\
0.5	0.118061830467837\\
0.55	0.112447248182878\\
0.6	0.107369296089614\\
0.65	0.102728090059553\\
0.7	0.0984885409552705\\
0.75	0.0945970448288429\\
0.8	0.0910148572873257\\
0.85	0.087707095157016\\
0.9	0.0846649117718081\\
0.95	0.0818389886105628\\
1	0.079180176627534\\
};
\addlegendentry{$ \LeakageOn = 0.5, \LeakageOff = 0.5$}

\end{axis}
\end{tikzpicture}%
    \caption{\textbf{Random telegraph input with dark current $\leakage$ and amplitude $\dynamicRange = 1$.} The information rate $\IR(\RTOn, \RTOff)$ is plotted for different values of $\leakage, \RTOn, \RTOff$. The graphs of $\RTOn = 0.1, \RTOff = 0.1$ and $\RTOn = 0.1, \RTOff = 0.5$ intersect, showing that dark current can alter the monotonicity properties of $\IR(\RTOn, \RTOff)$ in the $( \RTOn,\RTOff)$-plane.}
    \label{fig:Leakage}
\end{figure}
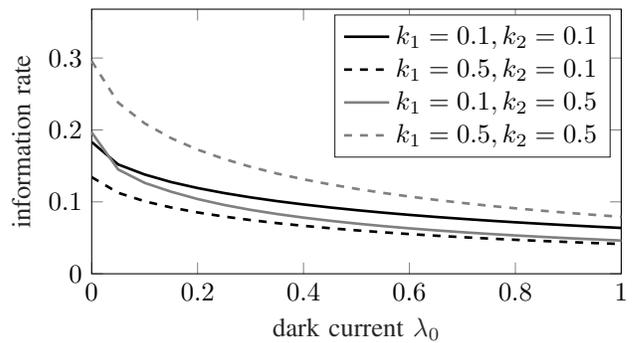

Results for different $\RTOn, \RTOff$ and increasing dark current are shown in figure \ref{fig:Leakage}. The information rate was computed with ODE system \eqref{eq:Density} - \eqref{eq:cummIR}. For any examined pair $(\LeakageOn, \LeakageOff)$, the information rate decreases with dark current as expected. The figure reveals a notable property. For fixed $\LeakageOn = 0.1$, the plots for $\LeakageOff = 0.1$ and $\LeakageOff = 0.5$ intersect. This means that increasing the dark current increases the information rate $\IR(0.1, 0.1)$ relative to $\IR(0.1, 0.5)$. Consequently, an increased dark current can qualitatively alter monotonicity and optimality properties in the $(\LeakageOn, \LeakageOff)$-phase plane. For example, the On favoring region increases with dark current \cite{Sinzger.2020}.

\subsection{Double On single Off}\label{cs:donsoff}
Binary Markov input processes exhaust the capacity when their switching rates tend to infinity. The defining property for optimality in the limit is only the proportion On/Off. With the autocorrelation time going to $0$ for an exhausting sequence of Semi-Markov processes also, the Markov property might as well be relaxed.
Consider the following capacity problem: We restrict the input process class to binary semi-Markov processes and impose a lower bound on the average sojourn times in the On and the Off state. Is the Markov case with its exponential sojourn times the capacity-achieving input? Here, we consider a Semi-Markov processes with exponential sojourn time in the Off and Erlang sojourn time in the On state. This can be realized as an instance of example class \ref{ex:Donsoff}, the afore-mentioned three-state Markov process that circles through one inactive and two active states. 
For the numerical evaluation we discretize $[0,1] \ni \state$. The state variables $( \Donsofffrac(\intime, \state), \z(\intime, \state))$ evolve according to \eqref{eq:Donsoff_frac} - \eqref{eq:Donsoff_cintensity}
with initial conditions $[ \Donsofffrac(0, \state), \z(0, \state)] = [\state, 1]$ and $\target(\intime, \state) = \Donsofffrac(\tau, \state)$. The times $\intime_\itime(\statefrom_\iPart)$ in \eqref{eq:matrix_version1}, that satisfy
\begin{equation}
\target(\intime, \statefrom_\iPart) = \intervalBound_{\ipart},
\label{eq:Donsoff_event}
\end{equation}
 were found by evolving the ODE system \eqref{eq:stationarity}, \eqref{eq:Donsoff_frac}, \eqref{eq:Donsoff_cintensity} and checking for the event \eqref{eq:Donsoff_event}. The matrix entries \eqref{eq:matrix_version1} were evaluated and $\idensity(\state)$ was found by fixed-point iteration with $2^\niter$ iterations. The rates $\DonsoffOn = \OnValue, \DonsoffR = \DonsoffOff = \OffValue$ exhibit a mutual information of $\IR(\msg, \out) = \DonsoffMIValue$ compared to $\IR(\msg, \out) = \RTMIValue$ for the Markov case.
 The rates were tuned such that the average sojourn times are the same. This example shows that the Markov input does not generally solve the capacity problem with average sojourn time constraint. It remains an open research question which On and Off sojourn time distributions are capacity-achieving.
 
The relevance for non-Markovian binary models is also reflected in biological systems. For instance, promoter models can exhibit multiple active and inactive states \cite{qiu.2019} while $\out_\ti$ counts mRNA synthesis events.
 
 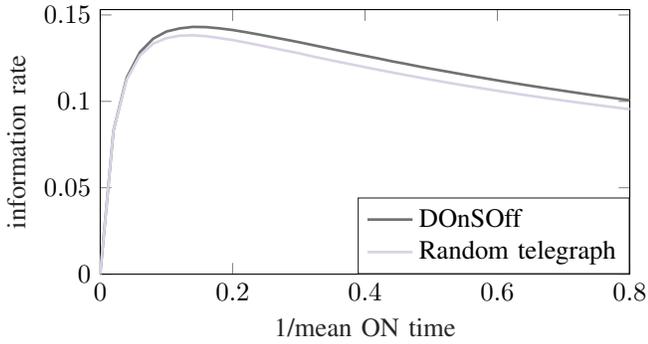
\begin{figure}
\centering
%
%
\definecolor{mycolor1}{rgb}{0.440000,0.44700,0.44100}%
\definecolor{mycolor2}{rgb}{0.85000,0.82500,0.89800}%
\begin{tikzpicture}

\begin{axis}[%
width=0.8*\the\myfigwidth,
height=0.4*\the\myfigwidth,
scale only axis,
xmin=0,
xmax=0.8,
xlabel style={font=\color{white!15!black}},
xlabel={1/mean ON time},
tick label style={/pgf/number format/fixed},
ymin=0,
ymax=0.153112700749594,
ylabel style={font=\color{white!15!black}},
ylabel={information rate},
axis background/.style={fill=white},
legend style={at = {(1,0)}, anchor = south east, legend cell align=left, align=left, draw=white!15!black}
]
\addplot [color=mycolor1, line width = 1pt]
  table[row sep=crcr]{%
0	0\\
0.02	0.0831031156691503\\
0.04	0.113747859588273\\
0.0600000000000001	0.128491503024752\\
0.08	0.136279012657972\\
0.1	0.140423231883193\\
0.12	0.142233336110019\\
0.14	0.143112700749594\\
0.16	0.142980459905931\\
0.18	0.14226727609368\\
0.2	0.141309173204423\\
0.22	0.140038487036053\\
0.24	0.138672981315829\\
0.28	0.135754744010959\\
0.3	0.134269923734162\\
0.38	0.127890199706857\\
0.42	0.124860435204725\\
0.44	0.123247797747414\\
0.46	0.121828640105825\\
0.48	0.120311859939686\\
0.5	0.118865651639192\\
0.52	0.117521793135546\\
0.56	0.114762264672848\\
0.6	0.112130957762987\\
0.64	0.109622018331585\\
0.68	0.1072484655843\\
0.74	0.103832477866943\\
0.8	0.100641649886509\\
};
\addlegendentry{DOnSOff}

\addplot [color=mycolor2, line width = 1pt]
  table[row sep=crcr]{%
0	0\\
0.02	0.0831118257542215\\
0.04	0.112637372702566\\
0.0600000000000001	0.126555791767222\\
0.08	0.13340369964893\\
0.1	0.136623824095287\\
0.12	0.138003672489067\\
0.14	0.138255082551559\\
0.16	0.137717384078399\\
0.18	0.136663030388682\\
0.2	0.135492152800386\\
0.22	0.134040925889672\\
0.24	0.132514471506289\\
0.26	0.130935802098112\\
0.3	0.127855197292901\\
0.32	0.126061167148137\\
0.34	0.124500948612553\\
0.36	0.122886627013608\\
0.42	0.118278206815997\\
0.46	0.115343981215648\\
0.5	0.112555485457832\\
0.54	0.109841780231285\\
0.6	0.106121162272212\\
0.64	0.103778226804121\\
0.68	0.10154895951414\\
0.72	0.0994264340584354\\
0.76	0.0973933672572479\\
0.8	0.0954561444112775\\
};
\addlegendentry{Random telegraph}

\end{axis}
\end{tikzpicture}%
\caption{\textbf{The Double On Single Off model} increases the mutual information rate compared to the random telegraph model. Parameters were $\DonsoffOn = \RTOn = \OnValue$ and $\DonsoffR = \DonsoffOff = 2\cdot \RTOff$. This choice guaranteed a model match in terms of mean On and Off times.}
\label{fig:Donsoff}
\end{figure}

\subsection{Hawkes process} \label{subsec:Hawkes}
The Hawkes process, introduced in \ref{ex:Hawkes}, is fully characterized in terms of second-order properties \cite{HAWKES.1971}, which was used in \cite{S.Shamai.1993} to obtain capacity upper bounds via the link to optimal linear estimation. Beyond information theoretic applications, the usefulness of the Hawkes process' ACID has been mentioned in the literature \cite{Oakes.1975,Daley.2003}. Oakes \cite{Oakes.1975} suggested that its equilibrium distribution, i.e., the ACID, can be found by iteratively solving an integral equation, but does not elaborate more on it. Daley \& Vere-Jones \cite[7.2.5 (iii)]{Daley.2003} provides a manual for setting up the integral equation and solving it with the method of steps. We are not sure whether Oakes meant this method by "iterative solution". We provide a different approach, using BReT-P and the fixed point iteration.
  The support of $\idensity(\state)$ is $(\HawkesBase + \HawkesJump, \infty)$ and contained in $(\HawkesBase, \infty)$. We consider the equidistant partition $\intervalBound_\ipart = \HawkesBase  + \ipart \Delta\state, \ipart = 0, \dots, \inum$ for $\Delta \state\cdot \inum$ large enough to cover most of the probability weight, i.e., for
\begin{equation}
    \int_0^{\Delta \state\cdot \inum} \idensity(\state + \HawkesBase + \HawkesJump) \dd \state \approx \int_0^{\infty} \idensity(\state + \HawkesBase + \HawkesJump) \dd \state.
\end{equation}
Then
\begin{equation}
    \intime(\stateto, \statefrom) = \HawkesDecay^{-1}\ln\left( \frac{\statefrom - \HawkesBase}{\stateto - \HawkesBase - \HawkesJump} \right)
\end{equation}
and
\begin{equation}
    \Density(\intime(\stateto, \statefrom), \statefrom) = \left(\frac{\stateto - \HawkesJump - \HawkesBase}{\statefrom - \HawkesBase} \right)^{\frac{\HawkesBase}{\HawkesDecay}} e^{-\HawkesDecay^{-1}(\statefrom - \stateto + \HawkesJump)}.
\end{equation}
Examples of the ACID, obtained from fixed point iteration as described in \ref{sub:Numeric_ACID}, are shown in fig. \ref{fig:Hawkes}. These may serve as initial distribution of $\CIntensity_0$ for the stationary Hawkes process. Using martingale theory, the equilibrium variance of $\CIntensity_\ti$ was derived (see Appendix \ref{Appendix:martingale}) to equal
\begin{equation}
   \V[\CIntensity_\infty] = \frac{\HawkesDecay\HawkesBase\HawkesJump^2}{2(\HawkesDecay - \HawkesJump)^2}.
   \label{eq:HawkesVariance}
\end{equation}
The parameter sets in fig. \ref{fig:Hawkes} were chosen, such that the ACID's first and second order moments are constant, but vary in the exponential decay parameter $\HawkesDecay$ of the Hawkes kernel. This makes the shape vary qualitatively. While for fast decay ($\HawkesDecay$ large) the region near the base value $\HawkesBase$ is frequented more heavily, for slow decay, the CI spends more time in the middle regime around the mean $\gain\HawkesMean$. This illustrates that the ACID analysis goes beyond the mean and variance analysis, i.e., that the ACID is parameterized by more than two parameters. 
\subsubsection{The Hawkes approximate marginal simulation}\label{cs:Hawkes filter}
For the purpose of comparing the Hawkes process to Markov-modulated Poisson processes and approximate filters, we view it as an approximate marginal simulation. It is obtained from the optimal linear filter approximation of the following process class. Let $\out_\ti$ be a doubly stochastic Poisson process whose external signal $\msg_\ti$ has the following first- and second-order statistics
\begin{equation}
    \E[\msg_\ti] \equiv \HawkesMean, \quad \Cov[\msg_\ti, \msg_s] = \HawkesVariance e^{-\HawkesAutocov \vert \ti - s \vert}.
    \label{eq:HawkesFirstSecond}
\end{equation}
And the $\sigmalg_\ti^{\msg, \out}$-intensity is $\intensityFct_\ti = \gain \msg_\ti$. External signals $\msg_\ti$ of this form comprise
(i) the random telegraph model with or without dark current and
(ii) the CIR process.
The optimal linear filter theory by Snyder \cite{Snyder.1991} identifies the estimator $\CIntensity_\ti$ that minimizes $(\tilde\intensity_\ti - \intensity_\ti)^2$ among all estimators of the following form, which is linear in $\out_{[0, \ti]}$,
$$\tilde\intensity_\ti = a(\ti) + \int_0^\ti h(\ti, u) \dd \out_u. $$
The resulting form of $\CIntensity_\ti$ appeals as a variant of the Kalman filter, with the Riccati equation for the innovation gain:
\begin{align}
    \dd \CIntensity_\ti = -\HawkesAutocov(\CIntensity_\ti - \HawkesMean) \dd \ti + \HawkesJump(\ti) (\dd \out_\ti - \CIntensity_\ti \dd \ti)\label{eq:Kalman}\\
    \frac{\dd}{\dd \ti} \HawkesJump(\ti) = -\HawkesJump(\ti)^2 - 2\HawkesAutocov \HawkesJump(\ti) + \frac{2\gain\HawkesAutocov\HawkesVariance}{\HawkesMean}. \nonumber
\end{align}
At equilibrium $\HawkesJump(\ti)$ can be replaced by the constant $\HawkesJump$ that solves $\HawkesJump^2 + 2\HawkesAutocov \HawkesJump - \frac{2\gain\HawkesAutocov\HawkesVariance}{\HawkesMean} = 0$. We interpret the evolution equation \eqref{eq:Kalman} as described in section \ref{ex:Gamma}. For the approximate marginal simulation $\out_\ti$ is self-exciting with CI $\CIntensity_\ti$ in contrast to $c\E[\msg_\ti\vert \sigmalg^\out_\ti]$ for estimation. So the eq. \eqref{eq:Kalman} is of the same shape as the Hawkes process.
A parameter match links the quadruple $(\HawkesMean, \HawkesVariance, \HawkesAutocov, \gain)$ to the original triple $(\HawkesBase, \HawkesJump, \HawkesDecay)$ as follows
\begin{equation}
    \HawkesJump = \sqrt{\HawkesAutocov^2 + \frac{2 \gain \HawkesAutocov\HawkesVariance}{\HawkesMean}} - \HawkesAutocov, \; \HawkesDecay = \HawkesAutocov + \HawkesJump,\; \HawkesDecay\HawkesBase = \gain\HawkesMean\HawkesAutocov.
    \label{eq:HawkesParameterLink}
\end{equation}
In the literature eq. \eqref{eq:HawkesFirstSecond} are also called the mean intensity and covariance density of the Hawkes process \cite{Gao.2018}.

We emphasize that the stationary distributions for common input processes, such as the CIR process, birth-death process or random telegraph model, are entirely characterized by mean and variance. The decay parameter $\HawkesAutocov$ in eq.~\eqref{eq:HawkesFirstSecond} is not captured by the stationary distribution. In contrast, the ACID does capture a change in $\HawkesAutocov$, see fig.~\ref{fig:Hawkes}, so it contains temporal information about the input process.

\begin{figure}
    \centering
    \input{Fig5_220110}
    \caption{\textbf{ACID for the Hawkes process.} ACID mean $\gain\HawkesMean = \frac{\HawkesDecay\HawkesBase}{\HawkesDecay - \HawkesJump} = 2$ and ACID variance $\frac{\HawkesDecay\HawkesBase\HawkesJump^2}{2(\HawkesDecay - \HawkesJump)^2} = 1$ were constant, while $\HawkesDecay \in \{0.3, 1, 3 \}$ varied. The truncation $\Delta \state\cdot \inum$ of the support was chosen to be the $0.999$-quantile of the Gamma distribution with mean $\gain\HawkesMean$ and variance $\gain^2\HawkesVariance$. Discretization granularity was $\inum = 200$. Additionally, three equidistant representatives $\state_\ipart$, see eq.~\eqref{eq:discretization} were chosen in each interval. Their mean function evaluations were used in eq.~\eqref{eq:matrix_version1} for the coefficients of $\OpI$. Number of iterations was $L = 15$.}
    \label{fig:Hawkes}
\end{figure}

We proceed with the second order analysis on the level of the output $\out_\ti$ in contrast to the input $\lambda_\ti$, and observe that for the process class in eq.~\eqref{eq:HawkesFirstSecond} the asymptotic mean and variance are known
$$\lim_{\ti \to \infty} \frac{1}{\ti} \E[\out_\ti] = \gain\HawkesMean, \quad \lim_{\ti \to \infty} \frac{1}{\ti} \V[\out_\ti] = \gain\HawkesMean + \frac{2\gain^2\HawkesVariance}{\HawkesAutocov}.$$
They are shared with the Hawkes process, see Appendix \ref{Appendix:martingale}. Hence, the exact and the approximate marginal simulation cannot be discriminated by first and second order analysis of $\out_\ti$. ACID is employed to detect the difference.
\subsubsection{Comparing the ACIDs of random telegraph with dark current and of the Hawkes process}\label{cs:Hawkes_comparison}
    \begin{figure*}
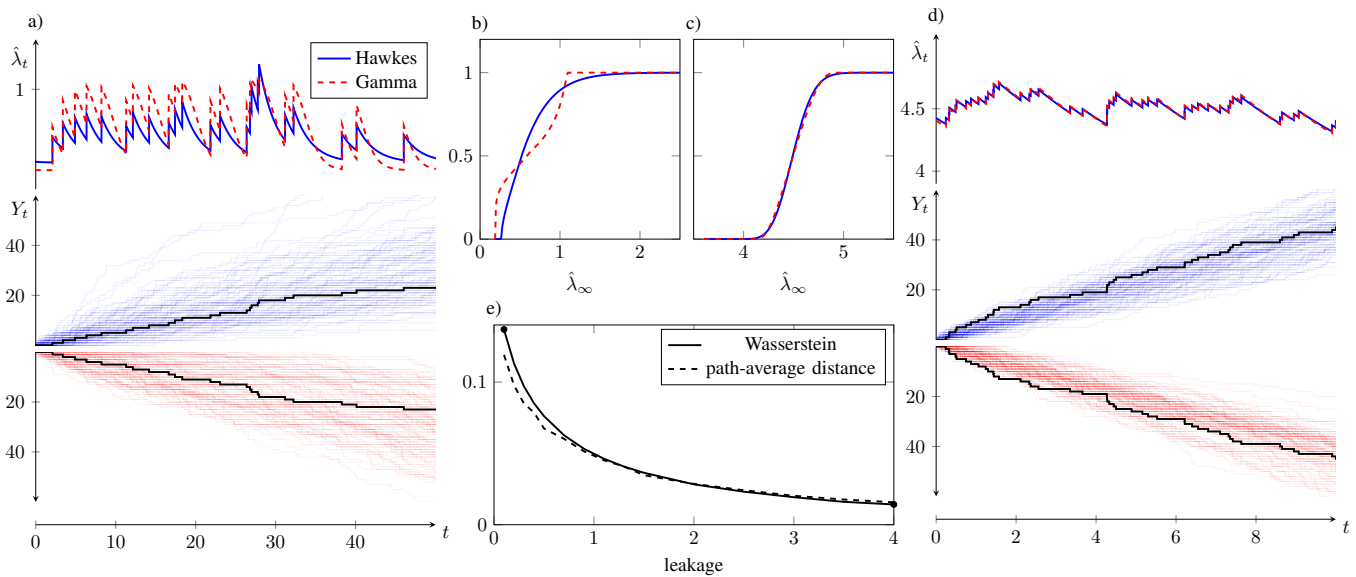

        \centering
        \resizebox{\textwidth}{!}{\definecolor{mycolor1}{rgb}{0.00000,0.044700,0.074100}%
\definecolor{mycolor2}{rgb}{0.085000,0.032500,0.09800}%
\begin{tikzpicture}

\begin{axis}[%
name = plot1b,
width=0.4\textwidth,
height=0.15\textwidth,
scale only axis,
x axis line style={draw=none},
xticklabel = \empty,
x tick style={draw=none},
xmin=0,
xmax=50,
ymin=0,
ymax=1.5,
ytick = {1},
ylabel = $\CIntensity_\ti$,
axis background/.style={fill=white},
axis lines = left,
axis x line*=top,
every axis x label/.style={at={(current axis.right of origin)},anchor=north west},
every axis y label/.style={at={(current axis.above origin)},anchor=north east},
yshift = 1mm
]
\input{fig7b_part1b_220211}
\end{axis}

\begin{axis}[%
width=0.4\textwidth,
height=0.15\textwidth,
name = plot1a,
at=(plot1b.below south west), anchor=above north west,
scale only axis,
axis lines = left,
x axis line style={draw=none},
xticklabel = \empty,
ylabel = $\out_\ti$,
x tick style={draw=none},
ytick = {20,40},
xmin=0,
xmax=50,
ymin=0,
ymax=60,
legend style={legend cell align=left, align=left, draw=white!15!black},
every axis y label/.style={at={(current axis.above origin)},anchor=north east},
axis background/.style={fill=white},
yshift = -1mm
]
\input{fig7b_part1a_220211}
\input{fig7b_part1_extra_220211}
\end{axis}

\begin{axis}[%
name = plot1c,
width=0.4\textwidth,
height=0.15\textwidth,
at=(plot1a.below south west), anchor=above north west,
scale only axis,
ytick = {20,40},
xmin=0,
xmax=50,
ymin=0,
ymax=60,
axis background/.style={fill=white},
axis lines = left,
x axis line style={draw=none},
xticklabel = \empty,
x tick style={draw=none},
y dir = reverse,
every axis y label/.style={at={(current axis.above origin)},anchor=north east},
yshift = 2mm
]
\input{fig7b_part1c_220211}
\input{fig7b_part1_extra_220211}
\end{axis}

\begin{axis}[%
name = plot1d,
width=0.4\textwidth,
height=0.01\textwidth,
at=(plot1c.below south west), anchor=above north west,
scale only axis,
xmin=0,
xtick={0, 10, 20, 30, 40},
xmax=50,
xlabel = $t$,
axis background/.style={fill=white},
axis lines = left,
y axis line style={draw=none},
yticklabel = \empty,
y tick style={draw=none},
every axis x label/.style={at={(current axis.right of origin)},anchor=north west},
yshift = 2mm
]
\addplot[color=black, opacity = 0, line width=0.2pt, forget plot]
  table[row sep=crcr]{%
0	0\\
10	0\\
};
\end{axis}

\input{fig7b_part2a_220214}
\input{fig7b_part2b_220214}
\input{fig7b_part2c_220214}

\begin{axis}[%
name = plot3b,
at=(plot2b.right of north east), anchor=left of north west,
width=0.4\textwidth,
height=0.15\textwidth,
scale only axis,
x axis line style={draw=none},
xticklabel = \empty,
x tick style={draw=none},
xmin=0,
xmax=10,
ylabel = $\CIntensity_\ti$,
ymin=3.9,
ymax=5.1,
ytick = {4,4.5},
axis background/.style={fill=white},
axis lines = left,
axis x line*=top,
every axis x label/.style={at={(current axis.right of origin)},anchor=north west},
every axis y label/.style={at={(current axis.above origin)},anchor=north east},
yshift = 1mm
]
\input{fig7b_part3b_220211}
\end{axis}

\begin{axis}[%
width=0.4\textwidth,
height=0.15\textwidth,
name = plot3a,
at=(plot3b.below south west), anchor=above north west,
scale only axis,
axis lines = left,
x axis line style={draw=none},
ylabel = $\out_\ti$,
xticklabel = \empty,
x tick style={draw=none},
ytick = {20,40},
xmin=0,
xmax=10,
ymin=0,
ymax=60,
legend style={legend cell align=left, align=left, draw=white!15!black},
every axis y label/.style={at={(current axis.above origin)},anchor=north east},
axis background/.style={fill=white},
yshift = -1mm
]
\input{fig7b_part3a_220211}
\input{fig7b_part3_extra_220211}
\end{axis}

\begin{axis}[%
name = plot3c,
width=0.4\textwidth,
height=0.15\textwidth,
at=(plot3a.below south west), anchor=above north west,
scale only axis,
ytick = {20,40},
xmin=0,
xmax=10,
ymin=0,
ymax=60,
axis background/.style={fill=white},
axis lines = left,
x axis line style={draw=none},
xticklabel = \empty,
x tick style={draw=none},
y dir = reverse,
every axis y label/.style={at={(current axis.above origin)},anchor=north east},
yshift = 2mm
]
\input{fig7b_part3c_220211}
\input{fig7b_part3_extra_220211}
\end{axis}

\begin{axis}[%
name = xaxis,
width=0.4\textwidth,
height=0.01\textwidth,
at=(plot3c.below south west), anchor=above north west,
scale only axis,
xmin=0,
xtick={0, 2,4,6,8},
xmax=10,
xlabel = $t$,
axis background/.style={fill=white},
axis lines = left,
y axis line style={draw=none},
yticklabel = \empty,
y tick style={draw=none},
every axis x label/.style={at={(current axis.right of origin)},anchor=north west},
yshift = 2mm
]
\addplot[color=black, opacity = 0, line width=0.2pt, forget plot]
  table[row sep=crcr]{%
0	0\\
10	0\\
};
\end{axis}
\node at (plot1b.above north west) {a)};
\node[anchor = south] at (plot2a.above north west) {b)};
\node[anchor = south] at (plot2b.above north west) {c)};
\node at (plot3b.above north west) {d)};
\node[anchor = south] at (plot2c.above north west) {e)};
\end{tikzpicture}}
        \caption{\textbf{The random telegraph with dark current modulates a Poisson process.} Comparison of the exact Snyder filter and the optimal linear filter (Hawkes process). a), d) Lower panel shows 100 realizations of the approximate marginal simulation (blue, increasing) and the exact simulation (red, decreasing). The lower y axis was flipped to simplify visual comparison. Upper panel compares $\CIntensity_\ti$ for the Hawkes (blue) and the exact (red, dashed). The trajectory $\out_\ti$ for which both where computed, is shown in the lower panel (black). b), c) The cdfs of the ACIDs are compared for Hawkes (blue) and Snyder filter (red, dashed). e) The Wasserstein metric between the Hawkes ACID and the exact ACID is depicted for increasing dark current. The dashed line depicts the path metric \eqref{eq:pathwise_metric} obtained from a) as the average distance between both paths. (The trajectory $\out_\ti$ was obtained from exact simulation.) All plots used the parameters $\dynamicRange = 1, \LeakageOn = \LeakageOff = 0.1, \gain = 1$. Dark current varied in e), while a), b) used $\leakage = 0.1$,  and c), d) used $\leakage = 4$. Both values are indicated as dots in e).}
       \label{fig:Wasserstein}
    \end{figure*}
For demonstration purposes we consider the tractable random telegraph input with dark current. It belongs to the considered input process class having exponentially decaying autocovariance function with $\HawkesAutocov = \LeakageOn + \LeakageOff$.
We compare the optimal linear filter (i.e. the Hawkes process) to the exact filter, obtained in \ref{ex:leakage}. As discussed at the end of the previous paragraph, a first and second analysis cannot reveal a difference. Also a visual inspection of a sample trajectory ensemble can hardly tell them apart (lower panels in fig.~\ref{fig:Wasserstein}a, d). And an asymptotic distribution for $\out_\ti$ cannot be compared because it does not exist. ACID is used to detect the parameter regimes where the approximate marginal simulation deviates from the exact marginal simulation (fig.~\ref{fig:Wasserstein}b, c). Figure \ref{fig:Wasserstein}e) shows that for fixed switching rates, gain and amplitude, the deviation gets more severe for smaller dark current.

While on the one hand two counting processes that agree in the path distribution share the ACID, on the other hand the same or similar ACID does not necessarily imply a similar path measure. ACID only gives the ensemble picture of the CI value at a typical time point after the process entered stationarity. To have a more path-wise comparison, the following metric is considered:
\begin{equation}
    \lim_{T \to \infty} \frac 1 T \int_0^T \vert\CIntensity_\ti - \CIntensity^H_\ti \vert \dd \ti .
    \label{eq:pathwise_metric}
\end{equation}
Here, $\CIntensity_\ti$ is the exact filter and $\CIntensity^H_\ti$ is the optimal linear filter, both evaluated as functions of the history $\out_{[0,\ti]}$ of a drawn sample path $\out_{[0,\infty]}$.
Ergodicity guarantees that the metric does not depend on the sample path. However, it is not clear whether to simulate the sample path from the exact process with CI $\CIntensity_\ti$ or the approximate process with CI $\CIntensity^H_\ti$. (For a sample trajectory obtained from $\CIntensity_\ti$ it is shown in fig.~\ref{fig:Wasserstein}e.) It depends on the context, i.e., the approximation goal, when to use a comparison of the ACIDs and when to use a path-wise comparison of the CIs for assessing an approximation. When approximating the information rate via eq.~\eqref{eq:Liptser} with an approximate filter $\CIntensity_\ti$, a comparison of the ACIDs seems suited.

\subsection{Gamma filter}\label{cs:Gamma}

The Gamma filter is used to illustrate the method for $\num = 2$. For technical details, we refer to Appendix~\ref{app:Wasserstein_Gamma}.

\subsubsection{Comparing the ACIDs of Gamma filter and Hawkes process}
Consider the Markov-modulated Poisson process $\out_\ti$ whose external signal is a CIR-process, i.e., satisfies eq. $\eqref{eq:HawkesFirstSecond}$. The Hawkes process seen as optimal linear filter and the Gamma filter both approximate the true CI of $\out_\ti$. We compare them for the same parameters $\HawkesMean, \HawkesVariance, \HawkesAutocov, \gain$.
First, we inspect the asymptotic mean slope and variance slope for the Gamma filter. By martingale techniques (see Appendix \ref{Appendix:martingale}), we can show that
\begin{equation}
    \lim_{\ti \to \infty} \frac{1}{\ti}\V[\out_\ti^G] = \gain\GammaMu + \frac{2\gain^2(\E[\GammaVar_{\infty}] + \V[\GammaMean_\infty])}{\GammaDecay }.
    \label{eq:FanoGamma}
\end{equation}
The Gamma filter satisfies the variance decomposition (see Appendix, A)
\begin{equation}
    \E[ \GammaVar_{\infty}] + \V[ \GammaMean_\infty] = \GammaSigma,
    \label{eq:VarianceDecomposition}
\end{equation}
consequently, the Gamma filter agrees with the exact CIR-modulated Poisson process in asymptotic first and second order moment. First and second order analysis cannot tell the Hawkes and Gamma filter apart.
A comparison of the ACIDs for a range of parameters reveals a slight increase of the Wasserstein metric for increased $\HawkesVariance$. The parameter $\HawkesAutocov$ had little effect on the Wasserstein metric. For technical details, see Appendix~\ref{app:Wasserstein_Gamma}.
Fig. \ref{fig:comparisonACID}b (Appendix) depicts the example with the largest Wasserstein metric among the considered parameters, revealing that the ACIDs are still very similar. Due to the ACID's limitation as partial characteristic, we cannot deduce that the path measures are close in some notion of distance.
However, we conclude the following. When quantities are computed that only depend on the ACID, the optimal linear filter - appealing with efficient analytic expressions - might replace the Gamma filter.

The mutual information rate along the Poisson channel was efficiently approximated by Monte Carlo simulation in \cite[Case study 1]{L.Duso.2019} via the Gamma filter. We replaced the Gamma filter by even the more efficient optimal linear filter.
 A comparison for this case is shown in Appendix, \ref{app:Duso}.
 Note, that both Gamma and Hawkes yield only an approximation of the exact information rate.
 Between them, the Hawkes can be preferred in this case with its gain in efficiency and no loss in accuracy relative to the Gamma.
In what respect the replacement works for more complicated reaction networks, must be carefully evaluated.

\subsection{Direct method for Hawkes and Dark Current}
\label{sub:direct}
In special cases, the fixed point method can be applied directly to the ACID $\acid(\aci)$, i.e., we derived a linear fixed point equation
\begin{equation}
    \acid(\aci) = \int \aciKernel(\aci, \acifrom)\acid(\acifrom) \dd \acifrom
    \label{eq:aciKernel}
\end{equation} 
for examples \ref{subsec:leakage} and \ref{subsec:Hawkes}.

\begin{theorem}
Let $\PDMP_\ti$ follow the stochastic evolution equation
\begin{equation}
    \dd \PDMP_\ti = \PDMPDyn(\PDMP_\ti) \dd \ti + [\PDMPTarget(\PDMP_\ti) - \PDMP_\ti] \dd\out_\ti
\end{equation}
and jumps of $\out_\ti$ occur with intensity $\intensityFct(\PDMP_\ti)$. Then for a differentiable initial condition $\PDMPDens(0, \PDMPState)$, the probability density evolves according to the PDE
\begin{align}
    &\partial_\ti \PDMPDens(\ti, \PDMPState) = \nonumber \\
    &-\partial_\PDMPState(\PDMPDyn(\PDMPState)\PDMPDens(\ti, \PDMPState)) - \intensityFct(\PDMPState)\PDMPDens(\ti, \PDMPState) + \PDMPTargInv'(\PDMPState) \intensityFct(\PDMPTargInv(\PDMPState)) \PDMPDens(\ti, \PDMPTargInv(\PDMPState)).
    \label{eq:Poisson-Liouville}
\end{align}
Suppose that the function $\PDMPHelp(\PDMPState)$ evolves as
\begin{equation}
    \PDMPHelp'(\PDMPState) = - \frac{\intensityFct(\PDMPState)}{\PDMPDyn(\PDMPState)}\PDMPHelp(\PDMPState)
\end{equation}
with arbitrary initial condition. If $\PDMPDens(\PDMPState)$ fulfills
\begin{equation}
    \PDMPDens(\PDMPState) = -\int_{\PDMPTargInv(\PDMPState)}^\UpperBound \frac{\intensityFct(\PDMPStatefrom) \PDMPHelp(\PDMPState)}{\PDMPDyn(\PDMPState)\PDMPHelp(\PDMPTarget(\PDMPStatefrom))} \PDMPDens(\PDMPStatefrom) \dd \PDMPStatefrom,
    \label{eq:DirectMethod}
\end{equation}
then the stationarity condition is satisfied:
\begin{equation}
    0 = -\partial_\PDMPState(\PDMPDyn(\PDMPState)\PDMPDens( \PDMPState)) - \intensityFct(\PDMPState)\PDMPDens(\PDMPState) + \PDMPTargInv'(\PDMPState) \intensityFct(\PDMPTargInv(\PDMPState)) \PDMPDens( \PDMPTargInv(\PDMPState)).
    \label{eq:station}
\end{equation}
\end{theorem}
\begin{proof}
Eq. \eqref{eq:Poisson-Liouville} is derived in the Appendix \ref{Appendix:Poisson-Liouville}. Eq. \eqref{eq:station} follows from \eqref{eq:DirectMethod} by Leibniz differentiation under the integral sign.
\end{proof}

The linear fixed point equation \eqref{eq:DirectMethod} can be used to return a numerical approximation of $\acid(\aci)$ directly. However, it can have a singularity at the equilibrium $\PDMPState$ characterized by $\PDMPDyn(\PDMPState) = 0$. And the ad-hoc discretization is not guaranteed to be a stochastic matrix as in eq.~\eqref{eq:matrix_version1}.

\begin{figure}
    \centering
    \input{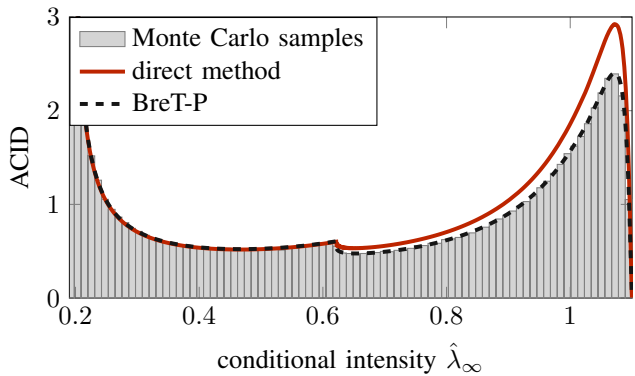}
    \caption{\textbf{Computation of the ACID with the direct method.} Parameters were as in fig.~\ref{fig:Leakage_MonteCarlo}. While the BReT-P agrees with the Monte Carlo simulation, the direct method, computed by fixed point iteration of \eqref{eq:direct_leakage}, shows an inaccuracy. The domain $(\RTRoot_1, \OnValue]$ was discretized into $N = 2000$ equidistant intervals.}
    \label{fig:direct_leakage}
\end{figure}

For the Hawkes process \ref{subsec:Hawkes} the equation \eqref{eq:DirectMethod} yields
\begin{equation}
    \aciKernel(\aci, \acifrom) = \frac{\acifrom(\aci - \HawkesBase)^{\frac{ \HawkesBase}{\HawkesDecay}-1}}{\HawkesDecay(\acifrom + \HawkesJump - \HawkesBase)^{-\frac{ \HawkesBase}{\HawkesDecay}}} e^{\frac{1}{\HawkesDecay} (\aci -\acifrom - \HawkesJump)}
    \label{eq:direct_Hawkes}
\end{equation}
in  \eqref{eq:aciKernel}. By "iterative solution" \cite[p.2]{Oakes.1975} this fixed point iteration could have originally been meant instead of the method of steps.
For \ref{subsec:leakage} note that eq.~\eqref{eq:Poisson-Liouville} corrects eq.\cite[section III.C]{Sinzger.2020} by a missing factor. The equation \eqref{eq:DirectMethod} yields
\begin{equation}
    \aciKernel(\aci, \acifrom) = \frac{\acifrom(\aci - \RTRoot_1)^{\frac{\RTRoot_1}{\Delta \RTRoot} - 1}(\RTRoot_2 - \PDMPTargInv(\acifrom))^\frac{\RTRoot_2}{\Delta \RTRoot}}{(\PDMPTargInv(\acifrom) - \RTRoot_1)^{\frac{\RTRoot_1}{\Delta \RTRoot}}(\RTRoot_2 - \aci)^{\frac{\RTRoot_2}{\Delta \RTRoot} + 1}}.
    \label{eq:direct_leakage}
\end{equation}
Fig.~\ref{fig:direct_Hawkes} (Appendix \ref{app:direct_method}) shows agreement for three Hawkes examples, while in fig.~\ref{fig:direct_leakage} the accuracy for the considered random telegraph with dark current suffers, possibly caused by the singularity at the equilibrium $\RTRoot_1$.

\section{Discussion}
The computation of the mutual information rate of a signal and its Poisson channel output via the expression due to Liptser remains a challenge. We contributed a new computational method by exploiting that the filtering distribution forms a piecewise-deterministic Markov process.
The Markovian nature of the filtering equation was also paraphrased as ‘mysterious’ concept of recursiveness by Brémaud \cite{Bremaud.1981}. Back in the days its purpose of saving memory-space was emphasized.
(A priori the CI depends on the history of $\out_\ti$. With the dependence on only the current state, there is no need to record the history.) We, in contrast, use the insight on the Markovian nature to analyze the asymptotic conditional intensity distribution.
We aimed at computing the ACID in a simulation-free way, i.e., without Monte Carlo simulations. This involved the evaluation of ODEs on a grid with the dimension given by the dimension of the sufficient statistic. We considered doubly stochastic Poisson processes (signal process along Poisson channel) and self-exciting counting processes. The ACID is accessible by our method in the case of very low number of non-zero signal states. With no limitations on the number of zero states, the method enables us to analyze binary semi-Markov inputs, i.e., have non-exponential phase-type Off-times. An interesting class of examples, that are characterized by low number of states of joint Markovian progression, are the approximate filters. By their very purpose of reducing the dimension of the sufficient statistic, they suit the limitations of our computational method.

The computational method is modular. The parametrization of the sufficient statistic $\state$ can be varied. The technique to find $\idensity$ according to eq.~\eqref{eq:version1} or \eqref{eq:version2} can be exchanged. The normalization constant can be employed from eq.~\eqref{eq:normalization_constant} or the resulting distribution $\density(\intime, \state)$ can be normalized. Depending on whether the entire ACID or summary statistics, such as variance or $\E[\phi(\cdot)]$, are of interest, the method can be modified. We expect that there is room for improvement: when substituting single modules in the method, the precision might be increased, computation time decreased and limitations relaxed. While our grid discretization was an ad-hoc approach, we consider our main contribution to lie in the formulation of BReT-P and the derivation of the integral boundary conditions.

We hope that the following semi-Markov perspective further guides improvements on the computational or theoretic side:
The parametrization $(\intime(\ti), \state(\ti))$ can be seen as the associated Markov process \cite[Chapter 3.4]{Limnios.2001} of the semi-Markov process $\state(\ti)$ with $\intime(\ti)$ being its backward recurrence time. Conditioned on being in state $\state$, the function $\cintensity(\intime, \state)$ is the hazard and $\Density(\intime, \state)$ has an interpretation as the survival function $\Prob[\wtime(\state) > \intime]$ of the sojourn time $\wtime(\state)$. The semi-Markov kernel is degenerate in $\state$ as
\begin{equation}
    \semiMarkovKernel(\statefrom, \stateto, \intime) = \Density(\intime, \statefrom)\Id(\target(\intime, \statefrom) = \stateto)
    \label{eq:semiMarkovKernel}
\end{equation}
because $\target$, the target anticipated at jumps, is a deterministic function. The embedded Markov chain has transition kernel
\begin{equation}
\tKernel(\statefrom, \stateSet) = \int_{\target(\intime, \statefrom) \in \stateSet} \cintensity(\intime, \statefrom)\Density(\intime, \statefrom) \dd \statefrom.
\end{equation}
The probability density of the embedded Markov chain (EMC) is proportional to $\idensity$ because it equals the probability density of being in a state $\state$ at a jump time. Consequently, eq.~\eqref{eq:version1} can be interpreted as the stationarity condition for the embedded chain. Finally, the relation $\density(\intime, \state) = \idensity(\state)\Density(\intime, \state)$ reflects the fact that the EMC's stationary distribution and the sojourn time factorize in the asymptotic \cite{Maes.2009}.

For the example \ref{cs:rand_tele} with zero-dimensional $\state$ the semi-Markov perspective boils down to a renewal process. Then eq. \eqref{eq:RT_survival_time} for $\Density$ expresses the known fact that the sojourn time for the marginal is a mixture of two exponential \cite{Kuczura.1973}.
Renewal processes in general fall within our framework with backward recurrence time $\intime(\ti)$ being sufficient without $\state(\ti)$. Their conditional intensity is $\CIntensity_\ti = -\frac{\dot\Density(\intime(\ti))}{\Density(\intime(\ti))}$ for the survival function $\Density(\intime)$. It is well established that the stationary distribution of $\intime$ has the pdf $\frac{\Density(\intime)}{\int \Density(u) \dd u}$ \cite{Cox.1962}. The transformation $\intime \mapsto -\frac{\dot\Density(\intime)}{\Density(\intime)}$ provides the ACID. This fraction is exactly the hazard.
    For the special case of exponential waiting times, the hazard rate is constant, so the ACID is a delta function with peak at the constant.
    
    The ACID can be interpreted as giving an ensemble perspective of the time-point-wise conditional intensity. Assuming that the counting process has reached the asymptotic regime, the ACID informs us how likely a certain intensity value is at a random time point.
    The ACID is a static quantity. Still, it captures temporal information to some extent. For instance, different autocovariance decays showed qualitatively different ACIDs for the Hawkes process. Most prominently, the ACID contains enough temporal information to inform the asymptotic rate of the path mutual information, as the Liptser expression shows.
    Limit results for the Hawkes process show that the variance function and covariance density are needed for the limit process in the central limit theorem \cite{Gao.2018}. These are not obtained from the ACID alone. The question, what properties of the counting process are uniquely determined by the ACID, remains open for future studies.
    
    Qualitatively, the ACID can exhibit non-smooth probability density functions. The ACID can serve as a partial characteristic for counting processes. In principal, it would be desirable to compare two counting processes in terms of their path measure. This can be too complex, in particular for self-exciting counting processes. The ACID may offer an accessible lower dimensional statistic for comparison that still exceeds first- and second-order analysis. We exemplified this for the comparison of an exact or approximate filter with an approximate filter. The ACID can discriminate between them if it differs. By our current state of knowledge, the ACID is limited as a partial characteristic of $\out_\ti$ in the opposite case. It does not allow a conclusion about the distance of the path measures in case two ACIDs are close. The usefulness of the ACID might be strengthened if statements can be found that allow conclusions of the following forms. (i) If $\lambda_\infty^{(1)}$ and $\lambda_\infty^{(2)}$ are close in distribution and the deterministic dynamics $F^{(1)}$ and $F^{(2)}$ are close, then the jump updates $f^{(1)}$ and $f^{(2)}$ are close. (ii) If $\lambda_\infty^{(1)}$ and $\lambda_\infty^{(2)}$ are close in distribution and $f^{(1)}$ and $f^{(2)}$ are close, then $F^{(1)}$ and $F^{(2)}$ are close. This is subject to future research and might allow to conclude the closeness of the path measures from the closeness of two ACIDs under mild additional conditions.
    The random time transform $\tilde{\out}_\to = \out(\int_0^\ti \CIntensity_s \dd s)$ is known to be a Poisson process. For instance it is used to assess goodness of fit for counting process models. In approximate marginal simulation, this approach can complement the ACID analysis. It also highlights the information content of integrals $\int_t^{t+h}\CIntensity_s \dd s$ for fixed gaps $h$. These could also be investigated asymptotically in distribution, fixing the ACIDs limitation of omitting the CI's temporal dependencies.
    
    Finally, we contributed to information theory by tackling the capacity problem for the Poisson channel with binary stationary input under average sojourn time constraints. The simulation-free computation employing BReT-P has the advantage to make partial derivatives easily accessible, which is useful in searching for optimal system parameters with gradient methods. We illustrated qualitative features of the Markov input without and with dark current. In addition, the case study of the double On single Off model showed, that the Markov input is not optimal. It remains an open research question what sojourn time distributions are optimal for the binary Semi-Markov input with average sojourn time constraints.

\appendices
\section{Reparametrization link}\label{app:repara}
The link between \eqref{eq:reparametrization} and \eqref{eq:integral_ISIT} can be obtained via the transformation rule applied on the transformation $\cprob\colon [0,\infty)\to (\RTRoot, 1]$. Let $T$ satisfy $\intime(\ti) \overset{\mathrm{d.}}{\to} T $, i.e., $T$ follows the stationary distribution of $\intime(\ti)$. Then
\begin{align*}
    \int_\RTRoot^1 \phi(z) \density_\condmean(z) \dd z &= \int_{\cprob^{-1}(\RTRoot)}^{\cprob^{-1}(1)} \phi(\cprob(\ti)) \density_\condmean(\cprob(\ti))\cdot \cprob'(\ti)\dd \ti \\&= \int_0^\infty \phi(\cprob(\ti)) \left[-\density_\condmean(\cprob(\ti))\cdot \cprob'(\ti)\right]\dd \ti \\&= \int_0^\infty \phi(\cprob(\ti)) \density_T(\ti) \dd \ti\\&= \idensity\int_0^\infty \phi(\cprob(\ti)) \Density(\ti) \dd \ti.
\end{align*}
\section{ODE system for partial derivatives}\label{app:derivative ODE system}
The joint evolution (for the time scaling $\intime \mapsto \gain \intime$, i.e., $\tilde f(\intime):=f(\intime/\gain)$ for $f = \density, \cprob, \density_1, \cprob_1, \cummIR_1$ and dropping the tilde again) is given by
\begin{align}
    \dot \density(\intime) &= - \cprob(\intime) \density(\intime) \label{eq:rho_evolution}\\
    \dot \cprob(\intime) &= \RTON - (\RTOFF + \RTON + 1) \cprob(\intime) + \cprob(\intime)^2\\
    \dot \density_1(\intime) &= \partial_1\dot \density(\intime) = -\cprob_1(\intime) \density(\intime) -\cprob(\intime) \density_1(\intime)\\
    \dot \cprob_1(\intime) &= \partial_1\dot \cprob(\intime) = 1 - \cprob(\intime) \nonumber\\& -(\RTOFF + \RTON + 1)\cprob_1(\intime) + 2\cprob(\intime)\cprob_1(\intime)\\
    \dot \cummIR_1(\intime) &= - \phi'(\cprob(\intime))\cprob_1(\intime)\density(\intime) - \phi(\cprob(\intime))\density_1(\intime) \label{eq:Jprime_evolution}
\end{align}
with initial conditions $(\frac{\gain \RTON}{\RTON + \RTOFF}, 1, \frac{\gain \RTOFF}{(\RTON + \RTOFF)^2}, 0, 0)$.
\section{Derivative with respect to gain}\label{app:partial3}
The scaling behaviour
\begin{equation}
    \IR(\RTOn, \RTOff, \gainSc\gain) = \gainSc\IR\left(\frac{\RTOn}{\gainSc}, \frac{\RTOff}{\gainSc}, \gain \right)
    \label{eq:gain_scaling}
\end{equation}
holds. Then for the derivative
\begin{align*}
   &\partial_3\IR(\RTOn, \RTOff, \gain)\\ &= \lim_{h\to 0} h^{-1} (\IR(\RTOn, \RTOff, \gain + h) - \IR(\RTOn, \RTOff, \gain))\\
   &= \lim_{h/\gain \to 0}\left(\frac{h}{\gain}\right)^{-1} \left(\IR\left(\frac{\RTOn}{\gain}, \frac{\RTOff}{\gain}, 1 + \frac{h}{\gain}\right) - \IR\left(\frac{\RTOn}{\gain}, \frac{\RTOff}{\gain},1\right)\right)\\
   &= \partial_3\IR\left(\frac{\RTOn}{\gain}, \frac{\RTOff}{\gain}, 1\right).
\end{align*}
\section{Variance of the ACID and asymptotic variance of $\out_\ti$}\label{Appendix:martingale}
The process equation for the Hawkes process is rewritten as
\begin{equation}
    \dd \CIntensity_\ti = -\HawkesAutocov(\CIntensity_\ti- \gain\HawkesMean)\dd \ti + \HawkesJump \dd \HawkesMartingale_\ti
\end{equation}
with the canonical $\sigmalg^\out_\ti$-martingale increment $\dd\HawkesMartingale_\ti = \dd \out_\ti - \CIntensity_\ti \dd \ti$. We are interested in the asymptotic behavior, i.e. $\CIntensity_\infty$. It can be brought to finite time $\ti$ under the shift \cite{Daley.2003} of the time domain $[0,\infty)$ to $(-\infty, t)$: 
\begin{equation}
    \CIntensity_\ti = \gain\HawkesMean + \int_{-\infty}^\ti e^{-\HawkesAutocov(\ti - s)} \HawkesJump \dd \HawkesMartingale_s.
    \label{eq:HawkesIntensityMartingale}
\end{equation}
By the Ito isometry for counting processes we get \eqref{eq:HawkesVariance} from \eqref{eq:HawkesIntensityMartingale} via:
\begin{align*}
    \V[\CIntensity_\ti] &= \E[(\CIntensity_\ti-\gain \HawkesMean)^2]\\
    &= \E\left[\left(\int_{-\infty}^\ti e^{-\HawkesAutocov(\ti - s)} \HawkesJump \dd \HawkesMartingale_s\right)^2 \right]\\
    &= \int_{-\infty}^\ti  e^{-2\HawkesAutocov(\ti - s)} \HawkesJump^2 \E[\CIntensity_s] \dd s\\
    &= \frac{\gain \HawkesJump^2 \mu}{2\HawkesAutocov} = \gain^2\HawkesVariance - \gain\HawkesMean\HawkesJump\\ &= \V[\intensityFct_\ti] - \E[\intensityFct_\ti]\HawkesJump.
\end{align*}
The equality $\HawkesJump$ that solves $\HawkesJump^2 + 2\HawkesAutocov \HawkesJump - \frac{2\gain\HawkesAutocov\HawkesVariance}{\HawkesMean} = 0$ yields \eqref{eq:HawkesVariance}.
Suppose we have an intensity process $\CIntensity_\ti = \gain \GammaMean_\ti$ that depends on some auxiliary process $\GammaVar_\ti$ via
\begin{equation}
    \dd \GammaMean_\ti = -\HawkesAutocov (\GammaMean_\ti - \HawkesMean) \dd \ti + \frac{\GammaVar_\ti}{\GammaMean_\ti}(\dd \out_\ti - \gain\GammaMean_\ti \dd \ti).
    \label{eq:HawkesGeneralization}
\end{equation}
The process $\GammaVar_\ti$ can be regarded as a conditional variance approximation.
For a process $(\GammaMean_\ti, \GammaVar_\ti)$
that satisfies \eqref{eq:HawkesGeneralization} we obtain
\begin{align}
  &\V[\out_T] = \E[(\out_T - \E\out_T)^2]\\ &= \E \left[ \left(\int_0^T c(\GammaMean_u - \GammaMu) \dd u + \int_0^T \dd \HawkesMartingale_\ti\right)^2\right]  \\
  &= \E \left[ \left(\int_0^T c \int_0^ue^{-\GammaDecay(u - \ti)} \frac{\GammaVar_\ti}{\GammaMean_\ti} \dd \HawkesMartingale_\ti \dd u + \int_0^T \dd \HawkesMartingale_\ti\right)^2\right]  \\
  &= \E \left[ \left(\int_0^T \frac{c}{\gamma} (1-e^{-\GammaDecay(T - \ti)})  \frac{\GammaVar_\ti}{\GammaMean_\ti} + 1 \dd \HawkesMartingale_\ti \right)^2\right] \\
  &=\E \left[ \int_0^T \left(\frac{c}{\gamma} (1-e^{-\GammaDecay(T - \ti)})  \frac{\GammaVar_\ti}{\GammaMean_\ti} + 1\right)^2 c\GammaMean_\ti \dd t \right].
  \label{eq:var_slope}
\end{align}
The chain rule yields the evolution equation of $\GammaMean_\ti^2$ from eq.~\eqref{eq:HawkesGeneralization}
\begin{align}
    \dd \GammaMean_\ti^2 &= \left\{-2\GammaDecay\GammaMean_\ti(\GammaMean_\ti - \GammaMu) - 2\gain\GammaMean_\ti \GammaVar_\ti \right\} \dd \ti \nonumber\\
    &+(\GammaMean_\ti + \frac{\GammaVar_\ti}{\GammaMean_\ti})^2 - \GammaMean_\ti^2 \dd \out_\ti.
    \label{eq:second_moment}
\end{align}
By applying the $\E$ and $\dd \out_\ti - c \GammaMean_\ti\dd \ti = \dd \HawkesMartingale_\ti$ we get
\begin{align*}
   \gain\E\left[\frac{\GammaVar_\ti^2}{\GammaMean_\ti}\right] &= \frac{\dd}{\dd \ti} \E[\GammaMean_\ti^2 ] + 2\GammaDecay \E[\GammaMean^2] - 2\GammaDecay\GammaMu^2\\
   &= \frac{\dd}{\dd \ti} \V[\GammaMean_\ti ] + 2\GammaDecay\V[\GammaMean_\ti ].
\end{align*}
Using this in eq.~\eqref{eq:var_slope}, we get
\begin{align}
\lim_{T \to \infty}\frac{1}{T}  \V[\out_T] &=\gain \GammaMu + \frac{2\gain^2}{\GammaDecay}(\E[\GammaVar_\infty] + \V[ \GammaMean_\infty]).
\label{eq:asymptotic_var_slope}
\end{align}
For the Gamma filter choose $\GammaVar_\ti$ as in eq.~\eqref{eq:Gamma_filter} and for the Hawkes choose $\GammaVar_\ti = \HawkesJump\CIntensity_\ti, \gain = 1$.

Asymptotic variance for the Gamma filter:
Taking the expectation in eq.~\eqref{eq:second_moment} and eq.~\eqref{eq:Gamma_filter} yields
\begin{align*}
   \frac{\dd}{\dd \ti}(\E[\GammaMean_\ti^2] -\GammaMu^2) &= \gain\E\left[\frac{\GammaVar_\ti^2}{\GammaMean_\ti}\right] -2\GammaDecay (\E[\GammaMean_\ti^2] - \GammaMu^2)\\
   \frac{\dd}{\dd \ti}\E[\GammaVar_\ti] &= -2\GammaDecay \E[\GammaMean_\ti^2] + 2\GammaDecay\GammaSigma -c\E\left[\frac{\GammaVar_\ti^2}{\GammaMean_\ti}\right].
\end{align*}
So the derivative of the sum evolves as \begin{align}
   \frac{\dd}{\dd \ti}( \E[\GammaVar_\ti] + \V[\GammaMean_\ti]) = -2\GammaDecay (\E[\GammaVar_\ti] + \V[\GammaMean_\ti]) + 2\GammaDecay\GammaSigma.
\end{align}
Since the sum starts in the steady state $\E[\GammaVar_0] + \V[\GammaMean_0] = \GammaSigma$, it stays constant for all $\ti$ and in particular in the asymptotic.

Both Gamma filter and Hawkes process satisfy the variance decomposition
\begin{equation}
    \E[\gain^2 \GammaVar_{\infty}] + \V[\gain \GammaMean_\infty] = \gain^2\GammaSigma, \quad \gain\HawkesMean\HawkesJump  + \V[\CIntensity_\infty] = \gain^2\HawkesVariance.
    \label{eq:VarianceDecomposition2}
\end{equation}
Consequently both filters agree with the process they approximate in terms of the asymptotic first and second order moments.

\section{Wasserstein metric}\label{app:Wasserstein_Gamma}
The Gamma ACID was computed using theorem \ref{theo:version1} and the method in section \ref{sub:Numeric_ACID}. The $(m, s)$-plane was truncated in a way to respect the minimal value of $m$ in the progression eq.~\eqref{eq:Gamma_filter} and to cover $99,5\%$ of the probability mass of a Gamma distribution with mean $\gain\GammaMu$ and $\gain^2\GammaSigma$. The bounds of the auxiliary $s$ were dictated by the minimum and maximum in eq.~\eqref{eq:Gamma_filter} for the above determined range of $m$. The rectangular $(m, s)$-domain was partitioned into $100\times 50$ congruent rectangles. Denote the boundaries of the rectangles by $\intervalBound^m_{\ipart}$ and $\intervalBound^s_{\ipart}$, respectively. Similar to the Hawkes ACID, $3\times 3$ equally spaced representatives $\state_\ipart$, see eq.~\eqref{eq:discretization}, were chosen in each rectangle. Their mean function evaluations were used in eq.~\eqref{eq:matrix_version1} for the coefficients of $\OpI$. As for the DOnSOff numerical approximation, the times $\intime_\itime(\statefrom_\iPart)$ in \eqref{eq:matrix_version1} that satisfy
\begin{equation}
\target_1(\intime, \statefrom_\iPart) = \intervalBound^m_{\ipart} \quad \text{or} \quad \target_2(\intime, \statefrom_\iPart) = \intervalBound^s_{\ipart}
\label{eq:Gamma_event}
\end{equation}
 were found by evolving the ODE system \eqref{eq:stationarity}, \eqref{eq:Gamma_filter} and checking for the event \eqref{eq:Gamma_event}.
 
 The Hawkes and Gamma filter were compared. A comparison of their ACIDs shows that they are remarkably similar. The cdfs of their ACIDs approximately agree for a range of values $\HawkesVariance, \HawkesAutocov$, while the mean and gain were kept constant $\HawkesMean = 2, \gain = 1$. The Wasserstein metric was computed for the regime $(\HawkesVariance, \HawkesAutocov) \in \{0.05, 0.1, \dot 2\} \times \{ 0.2, 0.4, \dots, 4 \}$. For the Gamma(Hawkes) filter, the numerical method yielded $93,0\%$ ($98,5\%$) of ACIDs that had mean value less than $0.01$ from the true value $\HawkesMean$ and $40,6\%$ ($53,1\%$) with a difference less than $0.001$. For the Gamma filter, most outliers (deviation $> 0.01$) were detected for $\HawkesAutocov = 0.05$ or $\HawkesVariance > 4\cdot\HawkesAutocov - 0.45$. For the Hawkes all outliers were detected at $\HawkesAutocov = 0.05$.
The Wasserstein metric values ranged from $0.0005$ to $0.13$. When neglecting $\HawkesAutocov = 0.05$ the largest value was $0.078$ for $\HawkesVariance = 4, \HawkesAutocov = 0.65$ and a decreasing trend for decreasing $\HawkesVariance$ was detected, relatively independent of $\HawkesAutocov$. The large values for $\HawkesAutocov = 0.05$ can be explained by the numerical inaccuracy in the Gamma ACID. Exemplary graphs for $\HawkesMean = 2, \HawkesAutocov = 0.65, \gain = 1$ and smaller vs. larger variance are depicted in fig. \ref{fig:comparisonACID}.
\begin{figure}
    \centering
    \input{Fig6_220116}
    \caption{\textbf{Comparison of the ACID for Gamma and Hawkes.} a) and b) show the cdf of the ACID for the Gamma filter and Hawkes process for $\HawkesMean = 2, \HawkesAutocov = 0.65$ and different values of $\HawkesVariance$. c) The Wasserstein metric was computed for a range of $\HawkesAutocov, \HawkesVariance$, while $\HawkesMean = 2, \gain = 1$ were fixed. In the examined regime $\HawkesAutocov$ has little effect, while the difference measure slightly increases with growing $\HawkesVariance$. The largest deviation for $\HawkesAutocov = 0.65, \HawkesVariance = 4$ was depicted in b) still showing agreement of the ACIDs.}
    \label{fig:comparisonACID}
\end{figure}
The most prominent dissimilarity is found in base values and it becomes more pronounced for larger variance. Still the ACIDs are very similar considering the example with the largest Wasserstein metric from the 
From this we conjecture, that when quantities are computed that only depend on the ACID, the optimal linear estimator might replace the Gamma filter. The mutual information rate along the Poisson channel was efficiently approximated by Monte Carlo simulation in \cite[Case study 1]{L.Duso.2019} via the Gamma filter. The Hawkes approximation can be conjectured to be even more efficient without severe loss in accuracy. In what respect the replacement works for this case and more complicated reaction networks, must be carefully evaluated.

\section{Computing the mutual information with Gamma filter vs. Hawkes}\label{app:Duso}
For the birth-death input process $\msg_\ti$ with birth rate $\HawkesAutocov\HawkesMean$ and death rate $\HawkesAutocov$ (i.e., mean $\HawkesMean$ and autocovariance function $\HawkesMean e^{-\HawkesAutocov \ti}$) the gamma filter's conditional variance equation is slightly modified:
\begin{equation}
   \dd\GammaVar_\ti = \{-\GammaDecay (2\GammaVar_\ti - \GammaMean_\ti - \GammaMu) - 2\gain\frac{\GammaVar_\ti^2}{\GammaMean_\ti} \} dt + \frac{\GammaVar_{\ti-}^2}{\GammaMean_{\ti-}^2}\dd \out_\ti.
        \label{eq:Gamma_filter_BD} 
\end{equation} 
In the limit $\gamma \to \infty$ ACID is a delta distribution at $\CIntensity_\infty = \E[\intensity_\infty]$ and hence
$$ \lim_{\gamma \ti \infty}\IR(\msg, \out) =  \E[\phi(\intensity_\infty)] - \phi(\E[\intensity_\infty]).$$
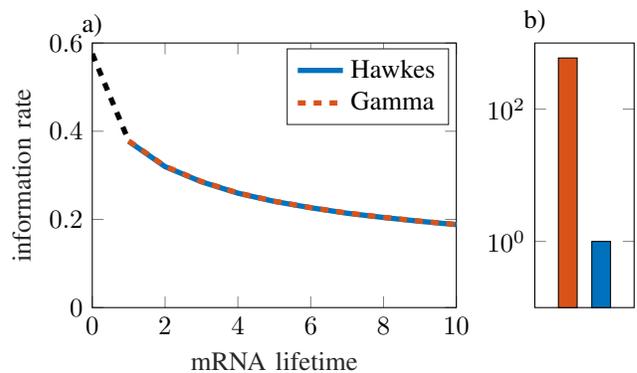
\begin{figure}
%
%
\definecolor{mycolor1}{rgb}{0.00000,0.44700,0.74100}%
\definecolor{mycolor2}{rgb}{0.85000,0.32500,0.09800}%
\begin{tikzpicture}

\begin{axis}[%
width=0.55*\the\myfigwidth,
height=0.8/2*\the\myfigwidth,
name = IR,
scale only axis,
xmin=0,
xmax=10,
xlabel style={font=\color{white!15!black}},
xlabel={mRNA lifetime},
ymin=0,
ymax=0.6,
ylabel style={font=\color{white!15!black}},
ylabel={information rate},
axis background/.style={fill=white},
legend style={legend cell align=left, align=left, draw=white!15!black}
]
\addplot [color=mycolor1, line width=2.0pt]
  table[row sep=crcr]{%
1	0.377055273178129\\
2	0.319637111557029\\
3	0.285242005197023\\
4	0.259461827106058\\
5	0.240848642980284\\
6	0.226441112958721\\
7	0.214009912570569\\
8	0.20435394134091\\
9	0.195750250255852\\
10	0.18819260993272\\
};
\addlegendentry{Hawkes}

\addplot [color=mycolor2, dashed, line width=2.0pt]
  table[row sep=crcr]{%
1	0.378025335823272\\
2	0.320455360451215\\
3	0.285886172846295\\
4	0.260002787811848\\
5	0.2412661303735\\
6	0.226855882361274\\
7	0.214393635493616\\
8	0.204665788600096\\
9	0.19609762928267\\
10	0.188539672265433\\
};
\addlegendentry{Gamma}

\addplot [color=black, dashed, line width=2.0pt, forget plot]
  table[row sep=crcr]{%
0	0.575632056716216\\
1	0.378025335823273\\
};
\end{axis}


\begin{axis}[%
width=0.15*\the\myfigwidth,
height=0.8*\the\myfigwidth/2,
at={(IR.right of north east)},
name = time,
anchor = left of north west,
scale only axis,
bar shift auto,
log origin=infty,
xmin=-0.2,
xmax=2.45,
xtick = \empty,
ymode=log,
ymin=0.1,
ymax=1000,
yminorticks=true,
axis background/.style={fill=white}
]
\addplot[ybar, bar width=0.5, fill=mycolor2, draw=black, area legend] table[row sep=crcr] {%
1	594.158203962373\\
};
\addplot[ybar, bar width=0.5, fill=mycolor1, draw=black, area legend] table[row sep=crcr] {%
1.25	1\\
};
\end{axis}
\node at (IR.above north west) {a)};
\node[anchor = south] at (time.above north west) {b)};
\end{tikzpicture}%
    \caption{Comparison of the information rate $\IR(\msg, \out)$ for a Markov-modulated Poisson process $\out$, approximately computed with the Gamma and Hawkes filter. The input $\msg$ was a birth-death process with birth rate $\gamma\mu$ and death rate $\gamma$. In the biological context this corresponds to the gene expression model with mRNA counts $\msg_\ti$ and protein translation event counts $\out_\ti$. a) The x-axis shows the average mRNA lifetime $\gamma^{-1}$. Both the Gamma and Hawkes approximate information rate were computed as Monte Carlo average using \cite[eq. (16)]{L.Duso.2019} with hyperparameters $T = 200$, sample size $10,000$. Parameters were $\mu = 10, c = 1$ and $\gamma = 1, \dots, 10$. The value at $0$ was determined analytically via $\E[\phi(\msg_\infty)] - \phi(\E[\msg_\infty])$, with $\msg_\infty \sim \Pois(\mu)$. b) shows the relative simulation time of the Gamma vs Hawkes in logarithmic scale with colors as in a)}. 
    \label{fig:comparisonDuso}
\end{figure}

\section{Derivation of the Liouville-Poisson master equation}\label{Appendix:Poisson-Liouville}
In Gardiner \cite{Gardiner.2009} the jump part of the differential Chapman-Kolmogorov equation requires a Kernel function $W(x \vert z,t)$ to act on the probability density, e.g. jumps are targeting a range of new values dictated by a probability kernel. In our case jumps target deterministic values, given by a function $f$.

Suppose $Z_t$ has a  trajectory-wise evolution
$$ dZ_t = A(Z_t) dt + [f(Z_t) - Z_t] dY_t $$
and jumps of $Y_t$ occur with intensity $\lambda(Z_t)$.

We derive the probability evolution equation. For this purpose let $f_{-}$ be the inverse of $f$, i.e. a jump that enters at $z$ jumped from $f_{-}(z)$.

It holds that
\begin{align*}
    &\Prob[Z(t+\Delta t)\in (-\infty, z]]\\ &= \int_{-\infty}^{f_{-}(z)} \Prob[\text{jump in } [t, t+\Delta t]\vert Z(t) = z'] p(z',t) \dd z' + o(\Delta t)\\
    &+ \int_{-\infty}^{z - A(z)\Delta t + o(\Delta t)} \Prob[\text{no jump in }[t, t+\Delta t] \vert Z(t) = z'] \times\\&\qquad \qquad \qquad \qquad \quad p(z',t) \dd z'\\
    &= \int_{-\infty}^{f_{-}(z)} \lambda(z')\Delta t p(z',t) \dd z' + o(\Delta t)\\
    &+ \int_{-\infty}^{z - A(z)\Delta t + o(\Delta t)} (1 - \lambda(z')\Delta t) p(z',t) \dd z'.
\end{align*}
Now we take the derivative with respect to $z$. This yields
\begin{align*}
    &p(z, t+ \Delta t)\\ = & f_{-}'(z) \lambda(f_{-}(z)) p(f_{-}(z),t) \Delta t\\
    &+ (1-\lambda(z)\Delta t)(1-A'(z)\Delta t)p(z-A(z)\Delta t,t) + o(\Delta t)\\
    = & f_{-}'(z) \lambda(f_{-}(z)) p(f_{-}(z),t) \Delta t\\
    &+ (1-\lambda(z)\Delta t)(1-A'(z)\Delta t)(p(z,t) - A(z)\Delta t p_z(z,t))\\ &+ o(\Delta t)\\
    = & f_{-}'(z) \lambda(f_{-}(z)) p(f_{-}(z),t) \Delta t + p(z,t)\\ &- \Delta t [\lambda(z)p(z,t) + A'(z)p(z,t) + A(z)p_z(z,t)] + o(\Delta t).
\end{align*}
Then
\begin{align*}
\lim_{\Delta t \to 0}&\frac{p(z, t+ \Delta t) - p(z, t)}{\Delta t}\\ =& - \partial_z(A(z)p(z,t)) - \lambda(z)p(z,t) \\&+ f_{-}'(z) \lambda(f_{-}(z)) p(f_{-}(z),t).
\end{align*}

\section{Comparison of BReT-P and direct method}\label{app:direct_method}
\begin{figure}
    \centering
    \input{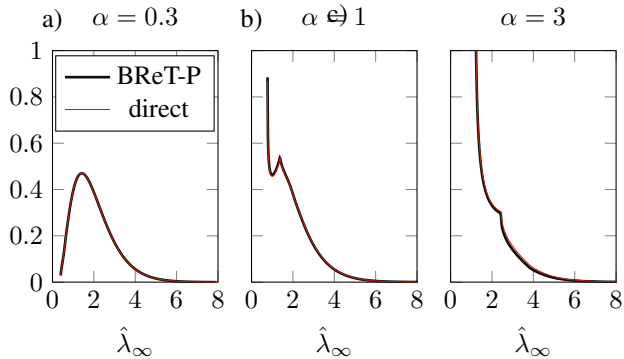}
    \caption{\textbf{Hawkes ACID: Comparison of BReT-P and direct method.} Parameters were as in \ref{fig:Hawkes}. Discretization granularity was $\inum = 1000$ for the direct method.}
    \label{fig:direct_Hawkes}
\end{figure}

Figure \ref{fig:direct_Hawkes} compares the BReT-P and the direct method and shows good agreement for the three cases.

  \section*{Acknowledgements}
    M.~Sinzger-D'Angelo would like to thank N.~Engelmann for discussion on the semi-Markov perspective, M.~Gehri for general discussion and feedback, as well as U.~Henning, S.~Startceva and B.~Alt for advice during the writing process.

\IEEEtriggeratref{40}
\bibliographystyle{IEEEtran}
\bibliography{2021_Transactions_ACID_file}

\end{document}